\documentclass[]{Manuscript}

\usepackage{enumitem}
\usepackage{color}

\lefttitle{ }
\righttitle{ }

\title{Capillary hysteresis induced by gap-resolved meniscus dynamics on Faraday instability in Hele-Shaw cells}

\author{Xingsheng Li\aff{1}, Jing Li\aff{1} \and Xiaochen Li\aff{2}}

\affiliation{\aff{1}Marine Numerical Experimental Center, State Key Laboratory of Ocean Engineering, School of Ocean and Civil Engineering, Shanghai Jiao Tong University, Shanghai, 200240, PR China
\aff{2}School of Civil Engineering and Transportation, South China University of Technology, Guangzhou, 510630, PR China}

\corresau{Jing Li, \email{lijing\_@sjtu.edu.cn}}

\begin{document}
\maketitle

\begin{abstract}

Existing theoretical analyses on Faraday instability in Hele-Shaw cells typically adopt gap-averaged governing equations and rely on Hamraoui's model coming from molecular kinetics theory, thereby oversimplifying essential transverse information, such as contact line velocity and capillary hysteresis, and conflicting with the unsteady meniscus dynamics.
In this paper, a gap-resolved approach is developed by directly modeling the transverse gap flow and the contact angle dynamics, which overcomes the aforementioned limitations, ultimately yielding a modified damping with respect to the static contact angle and hysteresis range.
A novel amplitude equation for linear Faraday instability is derived that combines this damping and the gap-averaged counterpart based on the oscillatory Stokes boundary layer, with the viscous dissipation preserved.
By means of Lyapunov's first method, an explicit analytical expression for the critical stability boundary is established.
Two series of laboratory experiments are performed that focus, respectively, on evolutions of the lateral meniscus and the longitudinal free surface near the Faraday onset, from which key parameters relevant to the theory are precisely measured.
Based on the experimental data, the validity of the proposed mathematical model for addressing the Faraday instability problem in Hele-Shaw cells is confirmed, and the generation and development mechanisms of the onset are clarified.
In the asymptotic analysis, the inclusion of contact angle dynamics increases the overall damping and thus partially compensates for the frequency detuning introduced by oscillatory Stokes flow approximation.

\end{abstract}


%{\bf MSC Codes }  {\it(Optional)} Please enter your MSC Codes here

\section{Introduction}
\label{sec:Introduction}

Faraday waves, a typical resonant phenomenon in parametric excitation systems, appear on the free surface of fluid when the container is subjected to periodic vertical vibration with sufficient amplitude and frequency.
Since first reported by \citet{faraday1831xvii}, advances in Faraday waves have proven to be fundamental for extensive scientific and industrial applications, including microscale assembly directed by liquid surface patterns \citep{chen2014microscale}, inkjet printing and deposition techniques \citep{arnold2007laser,turkoz2019reduction}, atomization \citep{liu2019experimental}, and sloshing mitigation in aerospace and marine systems \citep{faltinsen2017sloshing,colville2025faraday}.
These diverse applications have motivated numerous investigations into the underlying dynamics using experimental, theoretical, and numerical approaches \citep{benjamin1954stability,muller1993periodic,edwards1994patterns,kumar1994parametric,zhang1997pattern,perinet2009numerical,shao2021role}.

Recent experiments in Hele-Shaw cells have significantly enriched the knowledge of Faraday waves, revealing new insights into confined interfacial instabilities.
\citet{rajchenbach2011new} identified two novel classes of standing solitary waves of large amplitude, distinguished by odd and even symmetries.
Inspired by this pioneering work, a system of two coupled Faraday waves was observed at the two interfaces of three layers of fluids \citep{li2018observation}.
Furthermore, steady streaming patterns and self-organization of small tracers sustained by longitudinal Faraday waves were observed in narrow rectangular containers, though not strictly sorted into Hele-Shaw cells \citep{perinet2017streaming,alarcon2020faraday}.
In miscible fluids confined in similar narrow tanks of large aspect ratio, the Faraday instability can generate a turbulent mixing zone, as experimentally observed by \citet{briard2020turbulent} and numerically reported by \citet{grea2018final}.
These novel findings indicate that Faraday waves in tanks with a narrow gap exhibit more complicated dynamics than those in wide-mouth containers, with the governing mechanisms remaining incompletely understood.

Existing theoretical and numerical frameworks have focused predominantly on Faraday waves with spatially regular profiles, as observed by \citet{pradenas2017slanted} and \citet{li2019effect} in Hele-Shaw experiments.
What interests researchers is the critical condition for the emergence of these stationary patterns that uniformly occupy the entire laboratory-scale cell, which is called the Faraday onset.
Consistent with the parametric instability problem in wide-mouth containers \citep{benjamin1954stability,kumar1994parametric}, linear stability analysis serves as a common method for characterizing the onset conditions in Hele-Shaw configurations.
Current studies have pursued two distinct approaches: one is Floquet theory, which reduces to a generalized eigenvalue problem followed by a numerical solving process \citep{li2018faraday,bongarzone2023revised}; the other is Lyapunov's first method applied to the nonlinear dynamic system described by the amplitude equation, producing analytical expressions for the critical acceleration threshold \citep{rajchenbach2011new,li2019stability}.
The amplitude equation for the complex amplitude $A(t)$ has the form
\begin{equation}
\frac{\mathrm{d} A}{\mathrm{d} t}= \left ( \alpha _1+\mathrm{i}\alpha _2 \right ) A+\alpha _3A^*-\mathrm{i}\alpha _4\left | A \right | ^2 A,
\label{form of amplitude equation}
\end{equation}
where $\alpha_1$ characterizes system damping, $\alpha_2$ corresponds to the natural-response frequency detuning, $\alpha_3$ captures parametric forcing strength, and the last term is related to the nonlinear frequency shift with the wave amplitude.
Despite methodological differences, all stability analysis methods reveal the decisive dependence of critical threshold on the damping coefficient $\alpha_1$, whose precise quantification has been a persistent challenge.

One of the dominant dissipation originates from the no-slip condition at two lateral walls.
For sufficiently small gap sizes $b$ (typically satisfying $bk\ll1$ according to \citet{schwartz1986stability}, with $k$ being the wavenumber), the transverse flow is strongly constrained.
The inherent conflict between the no-slip boundary condition and vertical kinematics of Faraday waves makes it difficult to develop a fully three-dimensional mathematical model for this system.
Based on Darcy's law, \citet{gondret1997shear} developed an effective transverse averaging technique when studying the parallel flow driven by a given pressure gradient, which promotes the theoretical framework on the Faraday instability problem \citep{rajchenbach2011new}.
The assumption of a Poiseuille flow profile in the gap direction yields a parabolic velocity distribution that, when integrated and averaged, produces a characteristic damping coefficient $12\nu/b^2$, with $\nu$ denoting the kinematic viscosity.
Although \citet{rajchenbach2011new} incorporated this damping coefficient and proposed an amplitude equation that governs the dynamics of Faraday waves, their formulation is incomplete since the surface tension effects are entirely neglected.
\citet{li2018faraday} established the gap-averaged Navier--Stokes equations and implemented them in the open-source solver Gerris \citep{popinet2003gerris,popinet2009accurate}.
While the two-dimensional Faraday wave patterns were successfully reproduced and flow information was obtained, the lack of any contact angle model impedes accurate calculation of the instability threshold.
\citet{li2019stability} extended this theory by incorporating meniscus effects and modifying $\alpha_1$ with dynamic contact angle damping based on Hamraoui's model, subsequently deriving a new amplitude equation from the gap-averaged governing equations.
By means of Lyapunov's first method, they obtained an explicit expression for the instability threshold that shows substantial improvement due to the inclusion of contact angle dynamics.
However, systematic discrepancies between theory and experiment remain to be addressed.

Progress in this direction was recently made by \citet{bongarzone2023revised}.
Although Darcy's law accurately models unidirectional steady flows, its applicability to Faraday instability problem is limited by the inability to capture convective and unsteady inertial effects in periodically oscillating systems.
Inspired by the application of a classical Womersley velocity profile in pulsating flows in a channel \citep{womersley1955method,san2012improved} and the foundational work on sloshing dynamics in Hele-Shaw cells \citep{viola2017sloshing}, \citet{bongarzone2023revised} proposed a revised gap-averaged Floquet analysis.
The Poiseuille velocity profile was updated by the Womersley-like solution to Stokes boundary layer theory, resulting in a modified gap-averaged damping coefficient that accounts for oscillatory flow effects.
Contact angle damping was taken into account with Hamraoui's model employed by \citet{li2019stability}.
Comparison with experiments confirms that their predictions of the instability threshold are remarkably improved, revealing that Darcy's law indeed underestimates the effective gap-averaged damping.
Based on this oscillatory Stokes flow approximation, we have developed a fully viscous theory and found that bulk viscous dissipation plays a negligible role in linear stability analysis \citep{li2024stability}, while Hamraoui's model was maintained when dealing with the dynamic contact angle.
These analyses demonstrate that Stokes boundary layer theory yields a more accurate description of the gap flow.
Furthermore, this transverse flow approximation introduces a justifiable detuning, which was reported by \citet{bongarzone2023revised}, characterized by a shift in the response frequency, also manifested in the dispersion relation.
Through a series of experiments in a thin annular container, they confirmed how Darcy's law fails to capture this frequency detuning, which is essential for accurately predicting the locations of Faraday tongues.
This detuning, however, was not observed in rectangle Hele-Shaw cell experiments \citep{li2019stability}, suggesting that additional compensating effects may have been overlooked, and a possible incompatibility between Hamraoui's dynamic contact angle model and the oscillatory Stokes boundary layer may also exist.

The effect of dynamic contact angle, which has been neglected in many literature \citep{rajchenbach2011new,pradenas2017slanted,li2018faraday,rachik2023effects}, was first taken into account by \citet{li2019stability} when deriving the amplitude equation, embracing a model proposed by \citet{hamraoui2000can}:
\begin{equation}
\cos \theta =1-\frac{\beta }{\mu } Ca,\quad \text{with }Ca=\frac{\mu}{\sigma} w,
\label{Hamraoui model}
\end{equation}
where $\theta$ is the contact angle formed by the intersection of the free surface and the lateral wall at the contact line, $\beta$ is a friction coefficient, $\mu$ is dynamic viscosity, $\sigma$ is surface tension coefficient, and $Ca$ is the Capillary number defined by vertical velocity $w$.
Although its applicability has been verified in many investigations \citep{bongarzone2023revised,li2024stability}, there are several critical limitations.
First, this equation was originally developed for the kinetics of capillary rise, where the unidirectional flow feature guarantees mathematical rigor.
When applied to periodically oscillating systems,~\eqref{Hamraoui model} produces a nonphysical phenomenon ($\cos \theta >1$) for negative velocity ($w<0$).
In other words, this formulation contradicts the unsteady Stokes flow.
Second, existing theoretical analyses rely on the experimental values of $\beta$ measured by \citet{hamraoui2000can}, which were carried out in glass capillaries.
According to Young's equation, the wettability is sensitive to the solid substrate \citep{de1985wetting,good1992contact}, implying that $\beta$ necessarily depends on the container material characteristics, while previous work abused its measurements in different scenarios.
Finally,~\eqref{Hamraoui model} establishes a linear relationship between $\cos \theta$ and $w$.
The velocity $w$ refers to the motion of the contact line but is estimated by the vertical velocity of the free surface.
As will be demonstrated later in this paper, our laboratory observations indicate that before Faraday waves emerge, the contact line remains effectively pinned during initial excitation, while the contact angle varies within a specific range.
This widely used model cannot capture contact angle variations under pinned contact line conditions, namely contact angle hysteresis effects.

Based on numerous established theories on contact line boundary conditions in free surface flows \citep{hocking1987damping,cocciaro1993experimental,ting1995boundary,jiang2004contact,viola2018capillary,shao2021surface}, we aim at seeking a refined contact angle model that better captures the dynamics of meniscus.
Since the contact angle should be determined locally on the lateral walls, exhibiting a behavior dominated by transverse meniscus dynamics, its dimension is orthogonal to the in-plane Faraday waves.
Given the challenge of developing a complete three-dimensional mathematical model, we decompose the fluid domain into two distinct regions for separate analysis.
On the one hand, we focus on the Faraday wave profile and neglect the meniscus effects.
Following the approach taken by \citet{li2019stability}, we derive the amplitude equation for Faraday instability based on the oscillatory Stokes flow assumption.
Bulk viscous dissipation can be retained through the application of viscous flow theory.
Specifically, we will demonstrate why the contact angle dynamics cannot be incorporated directly into the gap-averaged governing equations.
On the other hand, we concentrate on the meniscus and try to directly resolve the transverse gap flow dynamics, which enables implementation of a more comprehensive contact angle hysteresis model and integration of contact angle damping into an amplitude equation.
The gap-resolved approach can preserve essential transverse information oversimplified by existing frameworks.
Then we verify the feasibility of incorporating the contact angle damping into the amplitude equation for Faraday instability.
Thus, an amplitude equation formed as~\eqref{form of amplitude equation} can be established, followed by a linear stability analysis.
Lastly, quantitative comparisons with experimental measurements are necessary to confirm the validity of the mathematical model.
Compared with recent progress in Faraday instability problem in Hele-Shaw cells that oversimplifies the transverse gap flow and obscures the true contact angle dynamics (e.g., \citet{li2019stability,bongarzone2023revised,li2024stability}), the proposed gap-resolved analysis represents a crucial contribution, as it successfully captures the meniscus dynamics near the onset, elucidates the role of contact angle hysteresis effects in the instability problem, and reveals the generation mechanism of Faraday onset.
The derived mathematical model overcomes the limitations of Hamraoui's linear dynamic contact angle model, whose predictions rely on the unpersuasive parameter $\beta$, and ultimately yields considerably improved agreement with experimental measurements of both the instability threshold and the critical wavenumber.
% In addition, compared with the amplitude equation proposed by \citet{li2019stability}, the present formulation offers three substantive refinements: (i) the gap-averaged damping is updated based on the oscillatory Stokes boundary layer; (ii) the contact angle dissipation is modified through the incorporation of the hysteresis model; and (iii) the bulk viscous dissipation is, for the first time, preserved explicitly via a rigorous mathematical derivation from the governing equations.}

The paper is organized as follows.
In \S~\ref{sec:Amplitude equation for Faraday waves}, we derive the amplitude equation for linear Faraday instability based on the Stokes boundary layer theory and retain the bulk viscous dissipation.
In \S~\ref{sec:Contact angle hysteresis damping}, we present a gap-resolved approach and derive the contact angle hysteresis damping.
The experimental methodology is described in \S~\ref{sec:Experiments}, together with some comparisons to numerically solved meniscus dynamics.
Linear stability analysis and theoretical model validation are performed in \S~\ref{sec:Results and discussion}, where other detailed discussions are also reported.
Finally, some conclusions are given in \S~\ref{sec:Concluding remarks}.

\section{Amplitude equation for linear Faraday instability}
\label{sec:Amplitude equation for Faraday waves}

\begin{figure}
\centerline{\includegraphics[width=13cm]{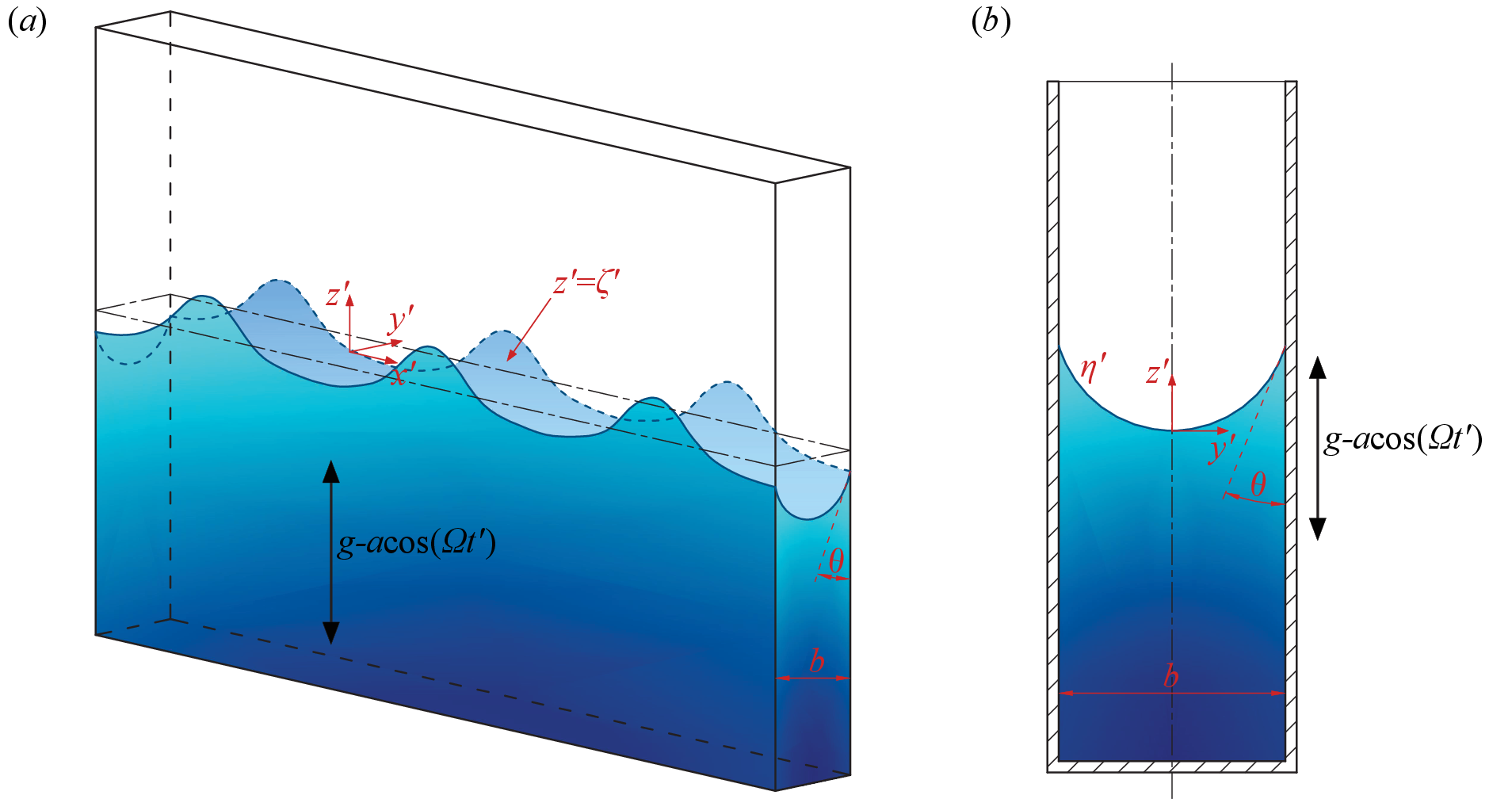}}
\caption{(\textit{a}) Sketch of Faraday waves in a Hele-Shaw cell that undergoes a vertical sinusoidal oscillation of acceleration amplitude $a$ and angular frequency $\Omega$.
The free surface elevation denoted by $\zeta' \left ( x',y',t' \right )$ contains both the two-dimensional Faraday wave profile along $x'$-direction and the meniscus in the gap direction.
(\textit{b}) View of the gap between two lateral walls.
Here $b$ denotes the gap size of the Hele-Shaw cell and $\theta$ is the contact angle of the liquid on the lateral walls.
The meniscus profile is denoted by $\eta'\left ( x',y',t' \right )$.}
\label{fig:sketch of faraday waves}
\end{figure}

We consider here a horizontally infinite Hele-Shaw cell of width $b$ filled with a kind of liquid of density $\rho$ and dynamic viscosity $\mu$ (kinematic viscosity $\nu = \mu / \rho$).
The container undergoes a vertical sinusoidal oscillation with acceleration amplitude $a$ and angular frequency $\Omega$.
In the Cartesian coordinate system as shown in figure~\ref{fig:sketch of faraday waves} that moves with the oscillating container, $z'=0$ is set at the bottom of the meniscus when at rest.
The central objective is to determine the critical condition for the onset of Faraday waves from an initially still free surface by stability analysis of the amplitude equation.

Recognizing that the linear Faraday instability is governed by the overall damping, which is an intrinsic property of the system, numerous studies have introduced a damping term into the Mathieu equation, based on which the stability analysis was carried out (e.g., \citet{kumar1994parametric,muller1997analytic,cavelier2022subcritical,bongarzone2023revised}).
Such approach has also been widely adopted in the context of amplitude equations: for example, \citet{milner1991square}, \citet{zhang1997pattern}, and \citet{rajchenbach2015faraday} modified the amplitude equation with damping terms and performed the weakly nonlinear analysis.
These studies confirm that adding or modifying the damping term in the amplitude equation is a rigorous way to improve the model fidelity.
Following this idea, the present analysis proceeds in two steps: first, by focusing solely on the Faraday wave in the $x'z'$ plane, we derive the amplitude equation for the linear Faraday instability problem in this section; then, by isolating the meniscus dynamics in the transverse gap direction, we seek a modification to the damping by taking contact angle hysteresis effect into account, which will be detailed in the next section.
The two orthogonal directions are connected through the constitution of the three-dimensional free surface as illustrated in figure~\ref{fig:sketch of faraday waves}, which is decomposed into two distinct components: the two-dimensional subharmonic Faraday wave profile $\bar{\zeta}'$ in the $x'z'$ plane, and the capillary driven harmonic meniscus wave $\eta'$ in the narrow gap direction.
The observed free surface deformation is a superposition of these two parts, namely
\begin{equation}
\zeta' \left ( x',y',t' \right ) =\eta' \left ( x',y' ,t'\right ) +\bar{\zeta}' \left ( x' ,t'\right ).
\label{free surface evolution}
\end{equation}
In typical derivations of the gap-averaged model (e.g., \citet{li2019stability,bongarzone2023revised}), taking account of the meniscus deformation introduces the out-of-plane curvature.
After gap-averaging, this curvature is linked to the contact line condition at lateral walls and eventually produces a contact angle damping term (refer to \S~\ref{sec:Incorporating meniscus effects into gap-averaged equations} for an explicit demonstration).
However, the gap-averaging treatment could oversimplify the meniscus dynamics, as will be elaborated in that section.

We derive a gap-averaged model for $\bar{\zeta}' \left ( x' ,t'\right )$
% Previously, \citet{li2024stability} have attempted to construct the stability analysis by gap-averaging the linearized three-dimensional Navier--Stokes equations directly.
% However, the complexity introduced by averaging both viscous governing equations and free surface boundary conditions was substantial, while the gain in accuracy proved marginal.
% We therefore adopt the alternative approach developed by \citet{bongarzone2023revised} gap-averaging the velocity field to represent the physics with respect to pressure.
with the following assumptions: (i) the three-dimensional Navier--Stokes equations and Taylor-expanded boundary conditions are linearized; (ii) the liquid depth is infinite, with a periodic condition in the $x'$-direction; (iii) the velocity component across the gap is zero, and the no-slip condition at lateral walls is employed; (iv) the oscillatory Stokes flow solution is derived retaining only the first subharmonic wave mode.
% ; (v) all governing equations and boundary conditions are gap-averaged; (vi) a multiple scale asymptotic analysis is performed, with the forcing and damping effects rescaled to enter at $\textit{O} (\epsilon)$.
An amplitude equation that contains both gap-averaged and linear viscous damping can be obtained, and will be modified to include an additional damping contribution due to contact angle hysteresis later on.

\subsection{Mathematical model for instability problem}
\label{sec:Mathematical model for instability problem}

\subsubsection{Oscillatory Stokes flow}
\label{sec:Oscillatory Stokes flow}

Stokes boundary layer theory has been effectively validated in characterizing the influence of two lateral walls in the Hele-Shaw flow \citep{viola2017sloshing, bongarzone2023revised, li2024stability}.
The primary procedure is to derive the velocity profile along the gap direction and then calculate its spatial average to quantify the damping induced by the lateral confinement.

Under vertical periodic vibration, the effective gravitational acceleration is given by $G\left ( t' \right )=g-a\cos \left ( \Omega t' \right ) $.
Linearizing the Navier--Stokes equations about the state of rest $\boldsymbol{U}'=0$ and $P'\left ( z',t' \right ) =-\rho G\left ( t' \right ) z'$, the equations for the perturbation fields of velocity, $\boldsymbol{u}'\left ( x',y',z',t' \right ) =\left ( u',v',w' \right )^\mathrm{T} $, and pressure, $p'\left ( x',y',z',t' \right )$, within the fluid bulk read
\begin{equation}
\frac{\partial \boldsymbol{u}'}{\partial t'}=-\frac{1}{\rho} \nabla' p'+\nu {\nabla'} ^2 \boldsymbol{u}',\quad
\nabla' \bcdot \boldsymbol{u}'=0.
\label{linearized N-S equations}
\end{equation}
According to the feature of Faraday waves in Hele-Shaw cells, the gap is narrow enough that $bk\ll 1$ with $k$ denoting the wavenumber.
We assume that there are no flows along the $y'$-direction and the velocity satisfies $v'=0$.
Equations in~\eqref{linearized N-S equations} then reduce to
\begin{subequations}
\begin{equation}
\frac{\partial u'}{\partial t'}=-\frac{1}{\rho} \frac{\partial p'}{\partial x'}+\nu \frac{\partial^2 u'}{\partial y'^2}, \quad 
\frac{\partial w'}{\partial t'}=-\frac{1}{\rho} \frac{\partial p'}{\partial x'}+\nu \frac{\partial^2 w'}{\partial y'^2}, \quad
\frac{\partial p'}{\partial y'}=0,
\label{dimensional NS equation}
\end{equation} 
\begin{equation}
\frac{\partial u'}{\partial x'}+\frac{\partial w'}{\partial z'}=0.
\label{dimensional continue equation}
\end{equation}
\end{subequations}
By making dimensionless with the following expressions:
\refstepcounter{equation}
\begin{equation}
\begin{aligned}
x=&\ kx',\quad y=\frac{y'}{b},\quad z=kz',\quad u=\frac{\Omega u'}{g},\quad v=\frac{\Omega  v'}{bkg},\\ &w=\frac{\Omega w'}{g},\quad p=\frac{k p'}{\rho g},\quad t=\Omega t',\quad \zeta=k\zeta',
\end{aligned}
\tag{\theequation{\textit{a}--\textit{i}}}
\label{dimensionless faraday waves}
\end{equation}
the first two equations in~\eqref{dimensional NS equation} in dimensionless form become
\refstepcounter{equation}
\begin{equation}
\frac{\partial u}{\partial t}=- \frac{\partial p}{\partial x}+\frac{\delta _{St}^2}{2} \frac{\partial ^2u}{\partial y^2},\quad \frac{\partial w}{\partial t}=- \frac{\partial p}{\partial z}+\frac{\delta _{St}^2}{2} \frac{\partial ^2w}{\partial y^2},
\tag{\theequation{\textit{a},\textit{b}}}
\label{two dimensional NS equation}
\end{equation}
where $\delta _{St}=\sqrt{2\nu/\Omega} /b$ is the dimensionless thickness of the oscillating Stokes boundary layer \citep{bongarzone2023revised}.

Because Faraday waves observed in Hele-Shaw cell experiments possess a subharmonic oscillation at a frequency of $\Omega/2$ \citep{li2019effect}, only the fundamental mode is considered, which is reflected in the separated exponential part of solutions associated with periodicity.
Therefore, the ansatzes can be written as $\boldsymbol{u}\left ( x,y,z,t \right )=\Real \left [ \tilde{\boldsymbol{u}} \left ( x,y,z\right ) e^{\mathrm{i}t/2 }\right ]$ and $p\left ( x,z,t \right )=\Real \left [ \tilde{p} \left ( x,z\right ) e^{\mathrm{i}t/2 }\right ]$.
Substituting them into~\eqref{two dimensional NS equation}, together with the no-slip condition at $y=\pm1/2$, the resulting ordinary differential equations about $y$ are solved.
Then we have
\refstepcounter{equation}
\begin{equation}
\tilde{u} =2\mathrm{i} \left \{ 1-\frac{\cosh \left [ \left ( 1+\mathrm{i} \right )y/\delta _1  \right ] }{\cosh \left [ \left ( 1+\mathrm{i} \right ) /2\delta _1 \right ] }  \right \} \frac{\partial \tilde{p} }{\partial x},\quad
\tilde{w} =2\mathrm{i} \left \{ 1-\frac{\cosh \left [ \left ( 1+\mathrm{i} \right )y/\delta _1  \right ] }{\cosh \left [ \left ( 1+\mathrm{i} \right ) /2\delta _1 \right ] }  \right \} \frac{\partial \tilde{p} }{\partial z},
\tag{\theequation{\textit{a},\textit{b}}}
\label{two dimensional velocity expression}
\end{equation}
where $\delta_1=\delta_{St}/\sqrt{1/2}$ is a modified dimensionless Stokes boundary layer thickness specific to the fundamental wave mode.

By integrating~\eqref{two dimensional velocity expression} along $y'$-direction and taking the average, the gap-averaged velocity components $\bar{u}$ and $\bar{w}$ are expressed by the two-dimensional pressure field $\bar{p}$, namely
\refstepcounter{equation}
\begin{equation}
\bar{u}\left ( x,z,t \right ) =\lambda_1 \frac{\partial \bar{p}}{\partial x},\quad
\bar{w}\left ( x,z,t \right ) =\lambda_1 \frac{\partial \bar{p}}{\partial z},
\tag{\theequation{\textit{a},\textit{b}}}
\label{gap-averaged velocity expression}
\end{equation}
with
\begin{equation}
\lambda_1=2\mathrm{i} \left \{ 1-\left ( 1-\mathrm{i} \right )\delta _1 \tanh\left [ \left ( 1+\mathrm{i} \right )/2\delta _1  \right ]   \right \}.
\label{expression of lambda}
\end{equation}
After this gap-averaging process, the Hele-Shaw flow reduces to two-dimensional dynamics in the $x'$-$z'$ plane and the dissipation resulting from the lateral walls is implicitly incorporated in the coefficient $\lambda_1$.

\subsubsection{Governing equations}
\label{sec:Governing equations}

The dimensionless continuity equation of~\eqref{dimensional continue equation}, when gap-averaged and combined with~\eqref{gap-averaged velocity expression}, is simplified to a Laplace equation for the pressure field:
\begin{equation}
\frac{\partial^2 \bar{p}}{\partial x^2}+\frac{\partial^2 \bar{p}}{\partial z^2} =0.
\label{laplace equation}
\end{equation}

The dimensionless non-penetrable boundary condition at the solid bottom reads
\begin{equation}
\frac{\partial \bar{p}}{\partial z}=0,\quad \text{at }z=-\infty.
\label{non penetrable condition}
\end{equation}
Assuming that the location of the bottom extends to $z=-\infty$ is feasible since the experimental liquid layer is sufficiently deep and $kH\gg1$ with $H$ denoting the liquid depth.

On the moving free surface, the normal stress is balanced by capillary force.
We mention here the Taylor expansion of boundary conditions on the free surface around $z'=0$ and, following eliminating the nonlinear terms, the normal stress boundary condition reads
\begin{equation}
-p'+\rho G\left ( t' \right ) \zeta'+2 \mu \frac{\partial w'}{\partial z'}=\sigma  \left ( \frac{\partial^2 \zeta'}{\partial x'^2} +\frac{\partial^2 \zeta'}{\partial y'^2}\right ),\quad \text{at }z'=0,
\label{dimensional normal stress condition}
\end{equation}
where $\sigma$ is the surface tension coefficient.
The right-hand side of~\eqref{dimensional normal stress condition} represents the capillary force, incorporating curvature contributions from two principal directions.
Because only the Faraday wave profile is considered in the current section (that is, only $\bar{\zeta}'$ in~\eqref{free surface evolution} is preserved), the meniscus curvature term $\partial^2 \zeta'/\partial y'^2$ in~\eqref{dimensional normal stress condition} should be removed.
After turning into dimensionless form using~\eqref{dimensionless faraday waves} and gap-averaging in the $y'$-direction, the boundary condition~\eqref{dimensional normal stress condition} becomes
\begin{equation}
-\bar{p}+\left ( 1- \frac{a}{g}\cos t \right ) \bar{\zeta}+\delta _{St}^2k^2b^2\lambda _1\frac{\partial^2 \bar{p}}{\partial z^2}=l_c^2 k^2\frac{\partial^2 \bar{\zeta}}{\partial x^2} ,\quad \text{at }z=0,
\label{dimensionless normal stress condition}
\end{equation}
where $l_c=\sqrt{\sigma /\rho g}$ denotes the capillary length, and $\bar{\zeta}$ represents the gap-averaged two-dimensional Faraday wave profile that has been made dimensionless with $\bar{\zeta}=k\bar{\zeta}'$.
In~\eqref{dimensionless normal stress condition}, the viscous stress reflected by the term $\partial^2 \bar{p}/\partial z^2$ is retained.
Although the previous theory indicates that the effect of fluid viscosity on dissipation in the Hele-Shaw system can be neglected \citep{li2024stability}, and this viscous term, scaling as $\delta _{St}^2k^2b^2\ll 1$, has been ignored by \citet{li2019stability} and \citet{bongarzone2023revised}, we retain it to preserve the fidelity of damping representation, with a detailed discussion on this viscous dissipation later.

The second boundary condition on the free surface is the linearized kinematic condition, whose dimensionless form is
\begin{equation}
\frac{\Omega^2}{gk}\frac{\partial \zeta}{\partial t}=w,\quad \text{at }z=0.
\label{Faraday kinematic boundary}
\end{equation}
After gap-averaging,~\eqref{Faraday kinematic boundary} reduces to
\begin{equation}
\frac{\Omega^2}{gk}\frac{\partial \bar{\zeta}}{\partial t}=\lambda_1\frac{\partial \bar{p}}{\partial z},\quad \text{at }z=0.
\label{averaged kinematic boundary}
\end{equation}

The system of equations~\eqref{laplace equation},~\eqref{non penetrable condition},~\eqref{dimensionless normal stress condition}, and~\eqref{averaged kinematic boundary} constitutes the governing equations for the Faraday instability problem in Hele-Shaw cells under the oscillatory Stokes flow approximation.

\subsection{Asymptotic expansion}
\label{sec:Faraday asymptotic expansion}

The derivation of amplitude equations is a classical method to describe pattern selection beyond linear instability (see, for instance, the frameworks of \citet{milner1991square}, \citet{chen1999amplitude}, and \citet{westra2003patterns}).
This approach is also applied to obtain the critical condition for linear Faraday instability in Hele-Shaw cells \citep{rajchenbach2011new,li2019stability}.
Inspired by these studies, we derive the amplitude equation from the reestablished mathematical model given in \S~\ref{sec:Governing equations}.

By introducing an expansion parameter $\epsilon$, the physical parameters can be expressed by
\refstepcounter{equation}
\begin{equation}
a=\epsilon \hat{a},\quad \delta _{St}^2k^2b^2=\epsilon \hat{\delta},\quad \lambda_1=2\mathrm{i}\left ( 1-\epsilon \hat{\delta}_1 \right ),
\tag{\theequation{\textit{a}--\textit{c}}}
\label{parameter expanding of Faraday}
\end{equation}
where $\epsilon \hat{\delta}_1=\left ( 1-\mathrm{i} \right )\delta _1 \tanh\left [ \left ( 1+\mathrm{i} \right )/2\delta _1  \right ]$ is the damping part of $\lambda_1$.
To examine the variation of variables in a long time scale, in the spirit of multiple scale analysis \citep{nayfeh1993introduction}, a slow time scale $T=\epsilon t$ is introduced, and naturally $\partial _t \to \partial _t+\epsilon \partial _T$.
Then the rescaled governing equations read
\begin{subequations}
\begin{equation}
\frac{\partial^2 \bar{p}}{\partial x^2} +\frac{\partial^2 \bar{p}}{\partial z^2}=0,
\label{rescaled laplace}
\end{equation} 
\begin{equation}
-\bar{p}+\left ( 1- \epsilon \frac{\hat{a}}{g}  \cos t \right ) \bar{\zeta}+ \epsilon \hat{\delta} 2\mathrm{i}\left ( 1-\epsilon \hat{\delta}_1 \right )\frac{\partial^2 \bar{p}}{\partial z^2} =l_c^2 k^2\frac{\partial^2 \bar{\zeta}}{\partial x^2},\quad \text{at }z=0,
\label{rescaled normal stress condition}
\end{equation}
\begin{equation}
4\omega ^2\left ( 1+l_c^2k^2 \right ) \left ( \frac{\partial \bar{\zeta}}{\partial t} +\epsilon \frac{\partial \bar{\zeta}}{\partial T} \right )=2\mathrm{i}\left ( 1-\epsilon \hat{\delta}_1 \right ) \frac{\partial \bar{p}}{\partial z},\quad \text{at }z=0,
\label{rescaled kinematic condition}
\end{equation}
\begin{equation}
\frac{\partial \bar{p}}{\partial z}=0,\quad \text{at }z=-\infty.
\label{rescaled bottom condition}
\end{equation}
\end{subequations}
In~\eqref{rescaled kinematic condition}, the angular frequency has been nondimensionalized with the expression $\Omega /2=\omega \sqrt{gk+\sigma k^3/\rho}$.
We shall define $\omega_0$ the limit of response frequency $\omega$ when $\epsilon=0$ with expansion $\omega^2 = \omega_0^2 + \epsilon \omega_1^2$.
The variables in~\eqref{rescaled laplace}--\eqref{rescaled bottom condition} are expanded as follows:
\refstepcounter{equation}
\begin{equation}
\bar{p}=\bar{p}_1 + \epsilon \bar{p}_2 + \textit{O}  ( \epsilon^2  ),\quad 
\bar{\zeta }=\bar{\zeta}_1 + \epsilon \bar{\zeta}_2 + \textit{O}  ( \epsilon^2  ).
\tag{\theequation{\textit{a},\textit{b}}}
\label{variables expanding of faraday}
\end{equation}
According to these definitions, we find that $\bar{p}\sim \textit{O} (1)$, $\bar{\zeta}\sim \textit{O} (1)$, $a \sim \textit{O} (\epsilon)$, $\delta _{St}^2k^2b^2 \sim \textit{O} (\epsilon)$, and the damping part of $\lambda_1$ is of the order $\sim \textit{O} (\epsilon)$.
In the absence of external forcing, the solutions corresponding to the static state are trivial, namely $\bar{\zeta}_s=0$ and $\bar{p}_s=0$.
Thus, the expansions in~\eqref{variables expanding of faraday} refer exclusively to the perturbed components.

The equations at each order are solved analytically; details are provided in Appendix~\ref{app:Asymptotic solutions of Faraday instability}.
At order $\epsilon$, a solvability condition arises from the Fredholm alternative, yielding
\begin{equation}
\frac{\mathrm{d} A}{\mathrm{d} T} +\frac{1}{2} \left [ \mathrm{i} \left ( \hat{\delta}_{1,r}+\mathrm{i} \hat{\delta}_{1,i} \right )+\mathrm{i}\omega_1^2+2\hat{\delta} \right ] A+\frac{\mathrm{i}\hat{a}}{4g\left ( 1+l_c^2k^2 \right ) }A^*=0,
\label{amplitude equation for faraday}
\end{equation}
where $\hat{\delta}_{1,r}$ and $\hat{\delta}_{1,i}$ are the real and imaginary parts of $\hat{\delta}_1$, respectively, $A(T)$ represents the complex amplitude of the linear Faraday wave on the slow time scale $T$, and the asterisk superscript ($*$) designates the complex conjugate.

Equation~\eqref{amplitude equation for faraday} is the amplitude equation for the Faraday instability problem that incorporates dissipation from both the confinement of the lateral walls and viscous stress, but excludes capillary damping induced by contact angle dynamics.
To derive an explicit expression of the damping, we decompose the complex amplitude $A$ into modulus and phase, namely $A=\left |A \right|e^{\mathrm{i}\Phi_A \left (T \right)}$, and write the modulus as $\left |A \right|=A_0 e^{-\gamma_1 t}$.
The damping rate $\gamma_1$ and phase $\Phi_A\left (T \right) $ are then given by
\begin{subequations}
\begin{equation}
\gamma_1 = -\frac{\mathrm{d} \log \left ( \left | A \right | /A_0 \right )  }{\mathrm{d} t} =-\frac{\epsilon }{\left | A \right | }\frac{\mathrm{d}\left | A \right | }{\mathrm{d}T} =\frac{1}{2} \left ( - \delta_{1,i} +2\delta_{St}^2k^2b^2 \right ),
\label{eq:gamma1}
\end{equation}
\begin{equation}
\frac{\mathrm{d} \Phi_A \left (T \right)}{\mathrm{d} T} =-\frac{1}{2} \left ( \hat{\delta}_{1,r} + \omega_1^2 \right ),
\label{eq:phase1}
\end{equation}
\end{subequations}
where $\delta_{1,r}$ and $\delta_{1,i}$ denote the real and imaginary parts of $\left ( 1-\mathrm{i} \right )\delta _1 \tanh\left [ \left ( 1+\mathrm{i} \right )/2\delta _1  \right ]$, respectively, and are associated with the gap-averaged damping.
The last term $\delta_{St}^2k^2b^2$ in~\eqref{eq:gamma1} reflects the viscous dissipation arising from the normal stress boundary condition~\eqref{dimensional normal stress condition}.
Equation~\eqref{eq:phase1} indicates the frequency shift introduced by the Stokes flow approximation.

\section{Contact angle hysteresis damping}
\label{sec:Contact angle hysteresis damping}

Existing theories for Faraday waves in Hele-Shaw cells typically employ the gap-averaging strategy.
Although the gap size $b$ is rather small and the assumption $kb \ll 1$ renders the instability problem nominally two-dimensional, the transverse meniscus determines the contact angle damping.
In this section, we first extend the analysis presented in \S~\ref{sec:Amplitude equation for Faraday waves} by incorporating meniscus effects, thereby demonstrating the limitation of Hamraoui's model~\eqref{Hamraoui model} and the infeasibility of implementing a contact angle model in the gap-averaged governing equations.
We then concentrate on the dynamics in the gap (see figure~\ref{fig:sketch of faraday waves}\textit{b}) and develop a more rigorous formulation for contact angle hysteresis damping.

\subsection{Incorporating meniscus effects into gap-averaged equations}
\label{sec:Incorporating meniscus effects into gap-averaged equations}

If we take the meniscus $\eta'$ into account and substitute~\eqref{free surface evolution} into~\eqref{dimensional normal stress condition}, another gap-averaged normal stress boundary condition is obtained that
\begin{equation}
-\bar{p}+\left ( 1- \frac{a}{g}\cos t \right ) \bar{\zeta}+\delta _{St}^2k^2b^2\lambda _1\frac{\partial^2 \bar{p}}{\partial z^2}=l_c^2 \left ( k^2\frac{\partial^2 \bar{\zeta}}{\partial x^2}+\frac{2}{b^2}\left. \frac{\partial \eta}{\partial y} \right |_{y=1/2} \right ),\quad \text{at }z=0,
\label{eq:gap dimensionless normal stress condition}
\end{equation}
where $\eta'$ has been nondimensionalized with $\eta=k\eta'$.
Comparing with~\eqref{dimensionless normal stress condition}, the last term in~\eqref{eq:gap dimensionless normal stress condition} refers to the curvature in the $y'$-direction, and is associated with the contact angle $\theta$ through the geometrical relation $\partial \zeta'/\partial y'|_{y'=\pm b/2}=\pm \cot \theta$, whose dimensionless form produces
\begin{equation}
\frac{\partial \eta}{\partial y}=\pm kb \cot \theta ,\quad \text{at }y=\pm \frac{1}{2}.
\label{geometrical relation of contact angle of Faraday}
\end{equation}

\begin{figure}
\centerline{\includegraphics[width=11cm]{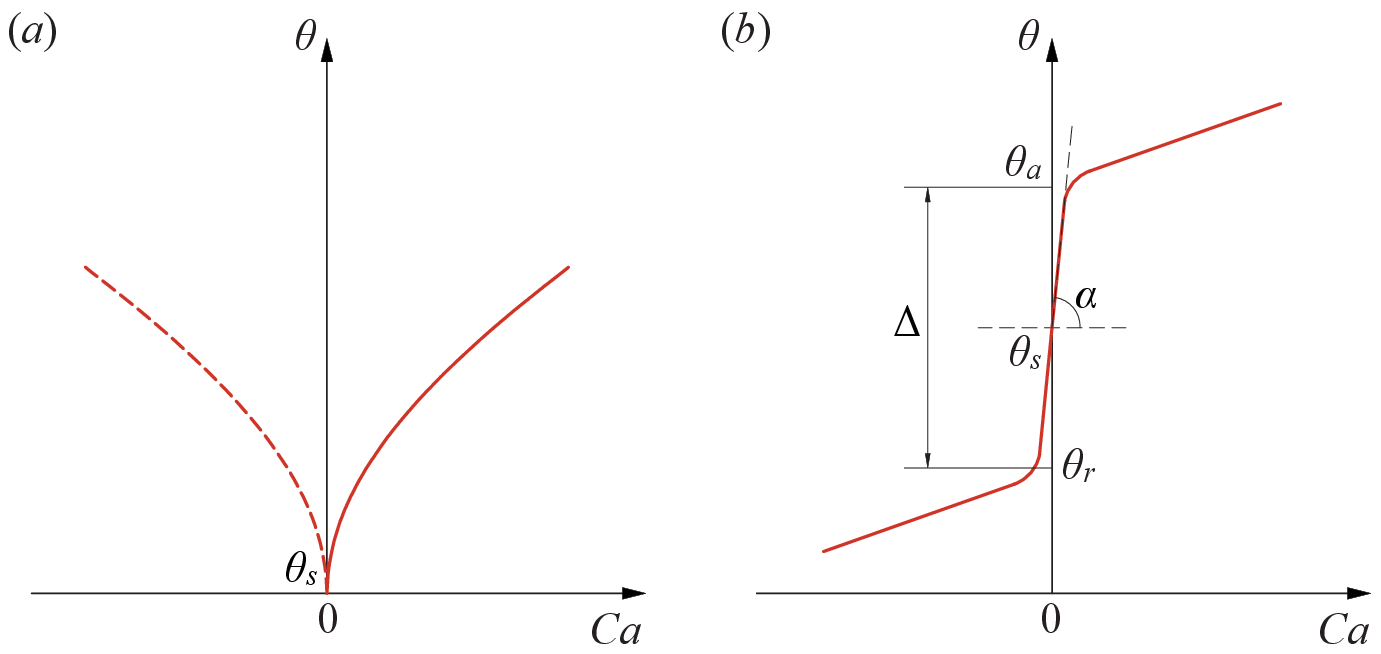}}
\caption{(\textit{a}) Sketch of the dynamic contact angle model~\eqref{Hamraoui model} developed by \citet{hamraoui2000can}.
(\textit{b}) Sketch of the contact angle hysteresis model~\eqref{hysteresis contact model} developed by \citet{viola2018capillary}.}
\label{fig:contact angle model}
\end{figure}

Before Faraday onset, the meniscus fluctuates up-and-down periodically in response to the forcing, and $\theta$ changes constantly around a static contact angle $\theta_s$.
Existing theoretical analyses typically use the dynamic contact angle model~\eqref{Hamraoui model} developed by \citet{hamraoui2000can} to define the variation of the dynamic contact angle with vertical velocity \citep{li2019stability,bongarzone2023revised,li2024stability}.
Figure~\ref{fig:contact angle model}\textit{a} shows the dependence of $\theta$ on $Ca$ as defined by~\eqref{Hamraoui model}, namely $\theta=\arccos{(1-\beta Ca/\mu)}$.
As mentioned previously, this model was originally formulated for capillary rise phenomena, which inherently requires $Ca\ge 0$ and thus only the right-hand part of the curve exists.
However, in oscillatory Hele-Shaw flows, the periodic nature permits negative velocity, namely $Ca<0$, leading to $\cos \theta>1$.
To avoid this nonphysical phenomenon, a typical approach is expanding around the static state and considering only the varying part \citep{bongarzone2023revised}, which yields
\begin{equation}
\frac{\partial \eta}{\partial y}=\mp\frac{kb\beta g}{\sigma \Omega} \bar{w},\quad \text{at }y=\pm\frac{1}{2}.
\label{Hamraoui boundary}
\end{equation}
The friction coefficient $\beta$ in~\eqref{Hamraoui boundary} is a phenomenological parameter that quantifies the dissipation rate resulting from the dynamic wetting process.
In fact,~\eqref{Hamraoui boundary} is the so-called Hocking's model \citep{hocking1987damping} describing a linear variation of the contact angle with the velocity of the contact line.
The two extremes $\beta=0$ and $\beta=\infty$ correspond to the free-end and pinned contact line, respectively.
For $\beta=0$, the contact line oscillates freely and the friction force does not work, which means the damping of the contact line is zero.
For $\beta=\infty$, the contact line is always fixed for whatever velocity.
This results in the damping approaching infinity and the instability will never emerge.
Both limiting cases are not consistent with the dynamics of Faraday instability.
In addition, the variation of the contact angle is not always proportional to the velocity in the present problem.
From laboratory observations, when forcing acceleration is far below the critical threshold, $\theta$ varies in a small range around $\theta_s$.
As the acceleration amplitude increases, this range expands.
But the contact line hardly moves before the onset threshold is reached.
This phenomenon coincides with the hysteresis feature of the dynamic contact angle, and cannot be captured by Hocking's model~\eqref{Hamraoui boundary}.
Once the linear Faraday instability appears, the free surface rises rapidly and, naturally, the contact line moves.

Inspired by \citet{viola2018capillary}, we introduce a refined contact angle model to deal with~\eqref{geometrical relation of contact angle of Faraday}, which reads
\begin{equation}
\theta =\theta_s+\frac{\rmDelta }{2} \tanh\left ( \alpha \varepsilon Ca  \right ),\quad \text{with }Ca=\frac{\mu}{\sigma}\frac{\Omega}{k} \frac{\partial \bar {\zeta}}{\partial t}.
\label{hysteresis contact model}
\end{equation}
Compared with the definition of $Ca$ in~\eqref{Hamraoui model}, the expression of $Ca$ in~\eqref{hysteresis contact model} is obtained by replacing the vertical velocity of the contact line with that of the gap-averaged free surface and combining with the kinematic condition~\eqref{averaged kinematic boundary}.
Figure~\ref{fig:contact angle model}\textit{b} illustrates how~\eqref{hysteresis contact model} represents the dynamic contact angle variation within the hysteresis range $\rmDelta$ around the static contact angle $\theta_s$.
The steepness parameter $\alpha \gg 1$ sets the velocity scale at which the contact angle varies.
The effect of pinned contact line is captured by the hyperbolic tangent function.
Although the contact line velocity is almost zero prior to Faraday onset, a constitutive relationship between $\theta$ and $Ca$ is required to model the contact angle dynamics.
The hyperbolic tangent function ensures rapid variation of $\theta$ at even small velocities and, as long as $\alpha$ is sufficiently large, the approximation in~\eqref{hysteresis contact model} remains consistent with the pinned contact line condition.
We note that figure~\ref{fig:contact angle model}\textit{b} gives a linear variation of $\theta$ beyond the hysteresis range, indicating a moving contact line.
This physical regime corresponds to the fully developed Faraday wave stage and is outside the scope of both~\eqref{hysteresis contact model} and linear stability analysis.

Because the fluid field is gap-averaged, one cannot distinguish the contact line motion near the lateral walls from the averaged interface elevation.
We utilize the vertical velocity of in-plane Faraday waves to define~\eqref{hysteresis contact model}.
Nevertheless, to bridge these two velocity scales, we introduce a dimensionless coupling coefficient $\varepsilon$, which may be treated to be either constant or time-dependent.
This process reflects how the gap-averaging method oversimplifies the transverse flow information.
In other words, the inclusion of meniscus effects in the gap-averaged normal stress boundary condition~\eqref{eq:gap dimensionless normal stress condition} inherently impedes the accurate modeling of contact angle dynamics.

Consequently, gap-averaged governing equations cannot properly account for contact angle effects, as they crudely approximate the contact line velocity through the gap-averaged vertical velocity of the free surface.
The widely used Hamraoui's linear model~\eqref{Hamraoui model} in existing gap-averaging frameworks fundamentally conflicts with the Faraday instability problem.
These constraints underscore the critical necessity of gap-resolved analysis in modeling realistic contact angle dynamics.

\subsection{Governing equations for the gap flow}
\label{sec:Governing equations for the gap flow}

When modeling the transverse gap flow, a distinct singularity arises: if the no-slip condition at the lateral walls is strictly enforced, the contact angle hysteresis model~\eqref{hysteresis contact model} that links the contact angle to the contact line velocity cannot be effectively introduced.
This inherent incompatibility indicates a fundamental challenge in formulating a full gap-resolved description of meniscus dynamics that simultaneously satisfies both the no-slip condition and a physically realistic contact angle model.
Fortunately, since both the gap-averaged damping arising from the no-slip condition at lateral walls and the linear viscous damping from the fluid bulk have already been included in the gap-averaged model of \S~\ref{sec:Amplitude equation for Faraday waves}, the objective here is solely to determine the contact angle hysteresis damping through the gap-resolved analysis of the meniscus wave.
To obtain this term, \citet{viola2018capillary} adopted the potential flow theory, which effectively circumvents the aforementioned singularity and significantly simplifies the mathematical model.
We follow a similar strategy and employ the potential flow theory to establish the gap-resolved model.
We emphasize that we do not seek to completely reconstruct the real three-dimensional flow field; hence, solutions in different directions are neither completely solved nor simply summed to obtain a full solution.
As elucidated at the beginning of \S~\ref{sec:Amplitude equation for Faraday waves}, focusing on the linear Faraday instability problem, the aim is to establish a more accurate amplitude equation and then apply with stability analysis to determine the critical condition for Faraday onset.
In such a linear framework, adding or modifying the damping term in the amplitude equation has been confirmed to be a rigorous approach.

Under vertical periodic vibration, equations that govern the dimensional velocity potential $\phi'$ and meniscus elevation $\eta'$ are summarized as follows:
\begin{subequations}
\begin{equation}
\frac{\partial^2 \phi'}{\partial x'^2} +\frac{\partial^2 \phi'}{\partial y'^2} +\frac{\partial^2 \phi'}{\partial z'^2} =0,
\label{eq:gap laplace equation}
\end{equation}
\begin{equation}
\frac{\partial \phi'}{\partial t'} + \left ( g-a\cos \Omega t' \right )\eta' = \frac{\sigma}{\rho} \left \{ \frac{\partial_{y'y'} \eta'}{\left [ 1+\left (\partial _{y'}\eta_s'  \right ) ^2 \right ]^{3/2} } + \frac{\partial_{x'x'} \eta'}{\left [ 1+\left (\partial _{y'}\eta_s'  \right ) ^2 \right ]^{1/2} } \right \},\quad \text{at }z'=0,
\label{eq:gap dynamic condition}
\end{equation}
\begin{equation}
\frac{\partial \eta'}{\partial t'} + \frac{\partial \eta_s'}{\partial y'} \frac{\partial \phi'}{\partial y'} = \frac{\partial \phi'}{\partial z'},\quad \text{at }z'=0,
\label{eq:gap kinematic condition}
\end{equation}
\begin{equation}
\frac{\partial \phi'}{\partial z'}=0,\quad \text{at }z'=-H,
\label{eq:gap bottom condition}
\end{equation}
\begin{equation}
\frac{\partial \phi'}{\partial y'}=0,\quad \text{at } y'=\pm \frac{b}{2},
\label{eq:gap lateral condition}
\end{equation}
\begin{equation}
\frac{\partial \eta'}{\partial y'}=\pm \cot\theta,\quad \text{at } y'=\pm \frac{b}{2},
\label{eq:gap geometrical relation}
\end{equation}
\begin{equation}
\theta=\theta_s + \frac{\rmDelta}{2}\tanh \left ( \alpha \frac{\mu}{\sigma} \left. \frac{\partial \eta'}{\partial t'} \right |_{y'=\pm b/2} \right ).
\label{eq:gap contact angle}
\end{equation}
\end{subequations}
Equations~\eqref{eq:gap dynamic condition} and~\eqref{eq:gap kinematic condition} are the dynamic and kinematic conditions on the linearized free surface, respectively.
Equations~\eqref{eq:gap bottom condition} and~\eqref{eq:gap lateral condition} represent the non-penetrable conditions at the solid bottom and lateral walls, respectively.
Because a numerical method will be used to solve the governing equations, the solid bottom is set at $z'=-H$.
This setting remains consistent with~\eqref{non penetrable condition}, since the liquid is deep enough that the distinction of the final results between finite and infinite values of $H$ is negligible.
Equations~\eqref{eq:gap geometrical relation} and~\eqref{eq:gap contact angle} work for the contact line condition.

Before solving the governing equations~\eqref{eq:gap laplace equation}--\eqref{eq:gap contact angle}, we decompose $\phi'$, $\eta'$, and $\theta$ into their static and perturbed components, namely
\refstepcounter{equation}
\begin{equation}
\phi' = \phi_s'+ \phi_p',\quad
\eta' = \eta_s'+ \eta_p',\quad
\theta = \theta_s+ \theta_p.
\tag{\theequation{\textit{a}--\textit{c}}}
\label{eq:gap dynamic and static}
\end{equation}
Unlike the perturbed quantities $\phi_p'$, $\eta_p'$, and $\theta_p$ that will be expanded in the asymptotic analysis, the static components $\phi_s'$, $\eta_s'$, and $\theta_s$ correspond to the equilibrium base state.
These steady terms remain decoupled from the perturbed parts and consequently do not influence the linear characteristics of~\eqref{eq:gap dynamic condition} and~\eqref{eq:gap kinematic condition}.
Similar treatment has been successfully implemented in modeling the capillary damping of inviscid surface waves in a circular cylindrical container \citep{kidambi2009capillary}.

Without external vibration, the static state satisfies
\begin{subequations}
\begin{equation}
\phi_s'=0,
\label{eq:phi_s=0}
\end{equation}
\begin{equation}
g\eta_s' = \frac{\sigma}{\rho} \frac{\partial _{y'y'}\eta_s' }{\left [ 1+\left (\partial _{y'}\eta_s'  \right ) ^2 \right ]^{3/2} },
\label{eq:eta_s}
\end{equation}
\begin{equation}
\frac{\partial \eta_s'}{\partial y'}=\pm \cot\theta_s,\quad \text{at } y'=\pm \frac{b}{2}.
\label{eq:theta_s}
\end{equation}
\end{subequations}
Equation~\eqref{eq:eta_s} is nonlinear in $\eta_s'$ and is solved numerically in MATLAB using an iterative Newton method \citep{viola2018capillary}.
The numerical procedure is detailed in Appendix~\ref{appen:Static meniscus}.

\subsection{Asymptotic expansion}
\label{sec:Gap asymptotic expansion}

By introducing an expansion parameter $\epsilon$, the physical parameters are expressed as $\rmDelta =\epsilon \hat{\rmDelta}$, $a=\epsilon \hat{a}$, and $\alpha \mu/\sigma =\hat{\alpha}/\epsilon$.
These physical quantities are rescaled in order to incorporate, as much as possible, all the features of the contact line law~\eqref{eq:gap contact angle}, for instance, a small hysteresis range and a rapid variation of contact angle at sufficiently small contact line velocity.

We present an asymptotic analysis of the perturbed components governed by~\eqref{eq:gap laplace equation}--\eqref{eq:gap contact angle}.
The variables are expanded as
\refstepcounter{equation}
\begin{equation}
\phi_p' = \phi_{p1}'+\epsilon \phi_{p2}'+ \textit{O} ( \epsilon ^2 ),\quad
\eta_p' = \eta_{p1}'+\epsilon \eta_{p2}'+ \textit{O} ( \epsilon ^2 ),\quad
\theta_p = \theta_{p1}+\epsilon \theta_{p2}+\textit{O} ( \epsilon ^2 ).
\tag{\theequation{\textit{a}--\textit{c}}}
\label{eq:gap variables expanding}
\end{equation}
The multiple scale approach \citep{nayfeh1993introduction} is also employed in this section.
Hence, we have a slow time scale $T'=\epsilon t'$ and $\partial _{t'} \to \partial _{t'}+\epsilon \partial _{T'}$.
Please note that the present asymptotic analysis follows a slightly different ordering from that of \citet{viola2018capillary}, where the static meniscus was expanded at the leading-order $\epsilon^0$.
% Instead, the expansions in~\eqref{eq:gap variables expanding} maintain the same asymptotic ordering used in~\eqref{variables expanding of faraday}, an approach supported by \citet{kidambi2009capillary}.}

\subsubsection{Order $\epsilon^0$}
\label{sec:Order epsilon^0}

At order $\epsilon^0$, the free-edge boundary condition is considered at the contact line, which means $\theta_{p1}=0$ and,~\eqref{eq:gap geometrical relation} reduces to
\begin{equation}
\frac{\partial \eta_{p1}'}{\partial y'}=0,\quad \text{at } y'=\pm \frac{b}{2}.
\label{eq:gap O1 contact line condition}
\end{equation}

To solve the equations at this order, the ansatzes are written as
\begin{subequations}
\begin{equation}
\phi_{p1}' \left ( x',y',z',t',T' \right )=B'\left ( T' \right ) \sin \left ( k_m x' \right ) \check{\phi}'_{p1} \left ( y',z' \right )e^{\mathrm{i}\Omega t'}+ \text{c.c.},
\label{eq:ansatz of phi1}
\end{equation}
\begin{equation}
\eta_{p1}' \left ( x',y',t',T' \right )=B'\left ( T' \right ) \sin \left ( k_m x' \right ) \check{\eta}'_{p1}\left ( y' \right )e^{\mathrm{i}\Omega t'}+\text{c.c.},
\label{eq:ansatz of eta1}
\end{equation}
\end{subequations}
where $B'(T')$ is the complex amplitude of the oscillation of the meniscus on the slow time scale.
It is essential to distinguish this amplitude used to describe the meniscus oscillation from the counterpart $A(T)$ of the in-plane subharmonic Faraday wave.
Existing studies have demonstrated that the meniscus wave oscillates harmonically \citep{shao2021surface,bongarzone2022subharmonic}, a behavior also confirmed by the subsequent experiments.
Hence, the harmonic temporal dependence is adopted in~\eqref{eq:ansatz of phi1} and~\eqref{eq:ansatz of eta1}.
The meniscus is assumed to vary along $x'$-direction sinusoidally with $\sin \left ( k_m x' \right )$.
During the numerical solving process at this order, the values of $k_m$ must be specified in advance.
Recalling that a traditional dispersion relation has been introduced at the same order in \S~\ref{sec:Faraday asymptotic expansion}, we adopt the results of $k_m$ calculated from this formulation.

Substituting~\eqref{eq:ansatz of phi1} and~\eqref{eq:ansatz of eta1} into the governing equations, we obtain a system of equations about the spatial variables $\check{\phi}'_{p1}$ and $\check{\eta}'_{p1}$, which are solved numerically by means of a spectral method \citep{viola2016mode,viola2018capillary}.
We refer to Appendix~\ref{appen:Numerical solutions at order epsilon0} for details on the numerical method and associated convergence analysis.

\subsubsection{Order $\epsilon$}
\label{sec:Order epsilon}

At order $\epsilon$, an amplitude equation that involves the contact angle hysteresis damping is derived.
The contact angle condition~\eqref{eq:gap contact angle} at this order reads
\begin{equation}
\theta_{p2}= \frac{\hat{\rmDelta}}{2}\tanh \left ( \frac{\hat{\alpha}}{\epsilon} \left. \frac{\partial \eta_{p1}'}{\partial t'} \right |_{y'= b/2} \right ).
\label{eq:gap O2 theta_2}
\end{equation}
The ansatzes for $\phi_{p2}'$ and $\eta_{p2}'$ are written as
\begin{subequations}
\begin{equation}
\phi_{p2}' \left ( x',y',z',t',T' \right ) =\sin \left ( k_m x' \right ) \tilde{\phi}_{p2}'\left ( y',z',T' \right )e^{\mathrm{i}\Omega t'} + \text{c.c.},
\label{eq:gap O2 ansatz of phi2}
\end{equation}
\begin{equation}
\eta_{p2}' \left ( x',y',t',T' \right ) =\sin \left ( k_m x' \right ) \tilde{\eta}_{p2}'\left ( y' ,T'\right )e^{\mathrm{i}\Omega t'} + \text{c.c.}
\label{eq:gap O2 ansatz of eta2}
\end{equation}
\end{subequations}

We introduce an adjoint global mode $ \boldsymbol{q}_1^\dagger =(\phi_{p1}^\dagger, \eta_{p1}^\dagger)^\mathrm{T} $ to obtain the solvability condition based on the Fredholm alternative.
By means of the Fourier expansion, the tangent function in~\eqref{eq:gap O2 theta_2} is properly treated to preserve hysteresis effects.
The detailed derivation of the solvability condition and treatment of~\eqref{eq:gap O2 theta_2} are given in Appendix~\ref{appen:Solvability condition for the gap flow}.
With these efforts, an amplitude equation is obtained, which reads
\begin{equation}
\frac{\mathrm{d} B'}{\mathrm{d} T'} + \frac{2\mathrm{i}\sin\theta_s\hat{\rmDelta } \sigma \kappa}{\pi \rho | \check{\eta}_{p1}' |_{y'=b/2} } \frac{B'}{| B' |} =0,\quad \text{with } \kappa =\frac{\left. \eta_{p1} ^{\dagger*} \check{\eta}_{p1}' \right |_{y'=b/2}}{ \int _{-b/2}^{b/2} \left. \left (  \eta_{p1} ^{\dagger*} \check{\phi}'_{p1} + \phi_{p1} ^{\dagger*} \check{\eta}'_{p1} \right )\right |_{z'=0} \mathrm{d} y'}.
\label{eq:gap amplitude equation}
\end{equation}
In accordance with the approach established in \S~\ref{sec:Amplitude equation for Faraday waves}, we nondimensionalize~\eqref{eq:gap amplitude equation} using $k^{-1}$ and $\Omega^{-1}$ as the characteristic length and time, respectively.
The resulting dimensionless amplitude equation reads
\begin{equation}
\frac{\mathrm{d} B}{\mathrm{d} T} + \chi B =0, \quad \text{with } \chi = \frac{2\mathrm{i}\sin\theta_s\hat{\rmDelta }\sigma \kappa}{\pi \rho \Omega | \check{\eta}_{p1}' |_{y'=b/2}|B'| }.
\label{eq:non-dimen gap amplitude equation}
\end{equation}

The amplitude equation~\eqref{eq:gap amplitude equation}, which exclusively accounts for the dissipation originating from contact angle dynamics, indicates that capillary hysteresis damping depends solely on the contact line velocity through its sign.
To isolate the damping contributions, following the approach used for~\eqref{amplitude equation for faraday}, we decompose the complex amplitude $B$ into modulus and phase, namely $B=\left | B \right | e^{\mathrm{i}\Phi_B(T)}$ with $\mathrm{d}\Phi_B(T)/ \mathrm{d} T=0$, and write the modulus as $\left |B \right|=B_0 e^{-\gamma_2 t}$.
Thus, the damping rate $\gamma_2$ is calculated by
\begin{equation}
\gamma_2 = -\frac{\mathrm{d} \log \left ( \left | B \right | /B_0 \right )  }{\mathrm{d} t} =-\frac{\epsilon }{\left | B \right | }\frac{\mathrm{d}\left | B \right | }{\mathrm{d}T} =\epsilon \chi= \frac{2\mathrm{i}\sin\theta_s \rmDelta \sigma \kappa}{\pi \rho \Omega | \check{\eta}_{p1}' |_{y'=b/2} \left | B' \right |}.
\label{eq:gamma2}
\end{equation}
Recalling that we have set $| \check{\eta}_{p1}' |_{y'=0} =1$, and combining with the definition of $B'(T')$ in~\eqref{eq:ansatz of phi1} and~\eqref{eq:ansatz of eta1}, $\left | B' \right |$ refers to the magnitude of the steady oscillation amplitude of the meniscus, determined right at the onset threshold, before the Faraday instability is fully established.
Although $|B'|$ is incorporated into the damping expression~\eqref{eq:gamma2}, the nonlinear character of the amplitude equations~\eqref{eq:gap amplitude equation} and~\eqref{eq:non-dimen gap amplitude equation} remains unchanged, indicating that the contact angle hysteresis damping depends on the amplitude of the oscillatory meniscus.
This fact agrees with the theoretical finding of \citet{viola2018capillary} when studying capillary hysteresis in sloshing dynamics and is consistent with the experimental observation by \citet{keulegan1959energy} and \citet{cocciaro1993experimental}.
\citet{viola2018capillary} further noticed that the influence of the capillary-induced force, which is associated with the contact angle hysteresis, depends only on the sign of the contact line velocity, thereby can be interpreted as a Coulomb-like friction force, as also pointed out by \citet{miles1967surface} in the context of damping of gravity waves.
Moreover, a similar mathematical formulation has been adopted by \citet{noblin2004vibrated} to model contact line dynamics in vibrated sessile drops.

We note some differences between the present problem and these earlier studies \citep{keulegan1959energy,cocciaro1993experimental,viola2018capillary}.
In those sloshing problems, an initial perturbation was applied to the fluid system, namely a non-zero initial value of $\left | B' \right |$; the subsequent wave attenuation was then measured with an amplitude that gradually decayed to zero.
In this restoring stillness process, both $\left | B' \right |$ and $\chi$ evolve continuously.
In contrast, the linear Faraday instability problem focuses on the critical condition for the onset of the free surface from a still state, a physically opposite process.
In the initial stage of Faraday instability, experiments (described below in \S~\ref{sec:Faraday onset}) show that the subharmonic Faraday wave amplitude grows slowly and exponentially, while the harmonic meniscus wave oscillates with a nearly constant amplitude.
Hence, the linear stability analysis examines the stability of the zero solution of the Faraday wave amplitude $A$, with the trivial solution expressed by $\left | A \right | = 0$ representing the still free surface.
In this stage, while the macroscopic Faraday waves have not yet emerged, the meniscus wave already oscillates at a harmonic frequency in response to the forcing, corresponding to a constant, non-zero amplitude $\left | B' \right |$.
This steady oscillation amplitude of the meniscus determines the contact angle hysteresis damping expressed by~\eqref{eq:gamma2} which is part of the dissipation hindering the Faraday onset.
Because $\left | B' \right |$ cannot be predicted a priori by the present theory, we employ experimentally measured values to calculate this gap-resolved capillary damping in the Faraday linear stability analysis.

\section{Experiments}
\label{sec:Experiments}

Currently, we have obtained explicit expressions of the dissipation induced by narrow transverse confinement, contact angle hysteresis effects, and fluid viscosity.
Comparison with experiments is required to verify the theoretical framework.
On the one hand, the evolution of the meniscus is observed, from which contact angle parameters such as the static contact angle $\theta_s$ and the hysteresis range $\rmDelta$ for three different liquids are obtained.
Comparison between numerical results of meniscus dynamics and experimental observations is also presented.
On the other hand, Faraday instability experiments are performed in Hele-Shaw cells made of the same material as the container used in meniscus experiments, with direct measurements of the onset parameters.

\begin{figure}
\centerline{\includegraphics[width=12cm]{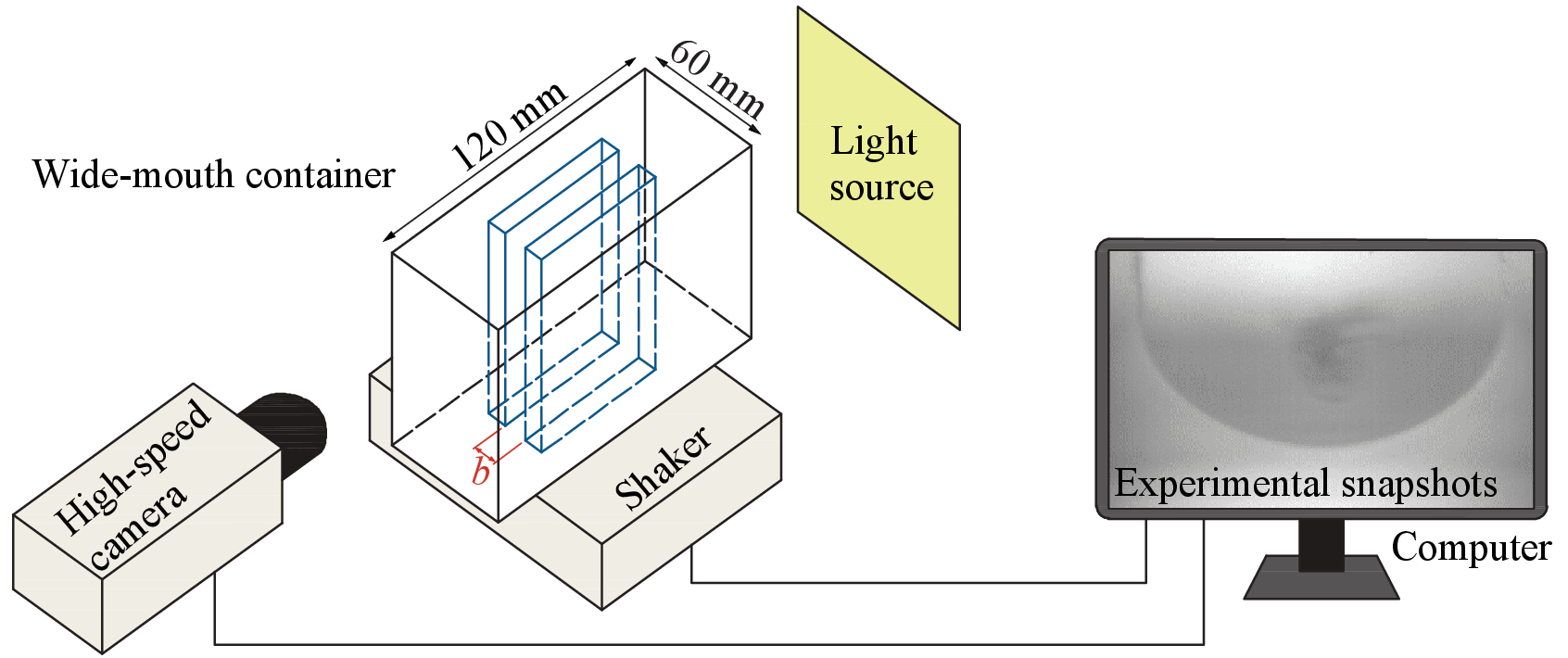}}
\caption{Schematic illustration of the experimental set-up for meniscus experiments, with the camera focused on the meniscus from a side view of the container.}
\label{fig:set-up_meniscus}
\end{figure}

The sketches of the experimental set-up are depicted in figure~\ref{fig:set-up_meniscus} and~\ref{fig:set-up_faraday}.
A container made of polyvinyl chloride (PVC) filled with liquids is fixed on an electrodynamic vibration generator (ESS-050), which can provide a peak force of 500 N and impose a vertical sinusoidal oscillation of acceleration $a\cos \Omega t'$, with $\Omega =2 \pi f$, and $f$ denoting the driving frequency.
A closed-loop system is utilized to control the vibration of the shaker through a computer, which guarantees a waveform deviation factor within $0.3\%$ and a driving frequency resolution of $0.01\%$.
The free surface is observed and recorded by a high-speed camera at a speed of 500 frames per second.
To ensure the illumination intensity and quality of images, a high-frequency lamp is used.
The temperature of the room is maintained at around $25\pm 1 \ ^\circ\mathrm{C}$.
Three ethanol--water mixtures are used as working liquids: pure ethanol ($>99.7\ \text{wt}\%$), $80\ \text{vol}\%$ and $70\ \text{vol}\%$ ($\pm 1 \ \text{vol}\%$) ethanol--water solutions.
Their physical properties are listed in table~\ref{tab:fluid parameters}.

\begin{table}
\begin{center}
\def~{\hphantom{0}}
\begin{tabular}{l@{\hspace{25pt}}c@{\hspace{25pt}}c@{\hspace{25pt}}c}
Liquid     & $\mu\ (\mathrm{mPa\ s})$ &   $\rho\ (\mathrm{g\ cm^{-3}})$ & $\sigma\ (\mathrm{mN\ cm^{-1}})$ \\[3pt]
99.7$\%$ pure ethanol             & 1.0995 & 0.7858 & 0.2207 \\
80$\%$ ethanol solution  & 2.0271 & 0.8529 & 0.2578 \\
70$\%$ ethanol solution  & 2.2249 & 0.8788 & 0.2735
\end{tabular}
\caption{Fluid parameters of the working liquids.
Data are taken from \citet{khattab2012density}.}
\label{tab:fluid parameters}
\end{center}
\end{table}

\subsection{Meniscus observation}
\label{sec:Meniscus observation}

In meniscus experiments, a wide-mouth container is employed (see figure~\ref{fig:set-up_meniscus} for the sketch), with a length of 120 mm, a width of 60 mm, and a height of 60 mm.
There are two panels made of PVC in the center of this container that form a 4.8 mm gap.
Therefore, the meniscus formed in this gap is consistent with that in Hele-Shaw cells.
The camera is placed parallel to these panels and focuses on the meniscus.
The Hele-Shaw cell is not employed as a working container in this experimental series because of the existence of two ends in the longitudinal direction of the cell.
Under the interaction between the liquid and the walls, the free surface is three-dimensional near the ends.
If we directly focus on the gap in Hele-Shaw cells, the liquid adhering to the end walls will disrupt the observation of a two-dimensional meniscus shape.
Using such a wide-mouth container, and optimizing focal length to isolate a specific cross-section, a two-dimensional free surface profile between the panels is captured and recorded by the high-speed camera.

\begin{figure}
\centerline{\includegraphics[width=13.5cm]{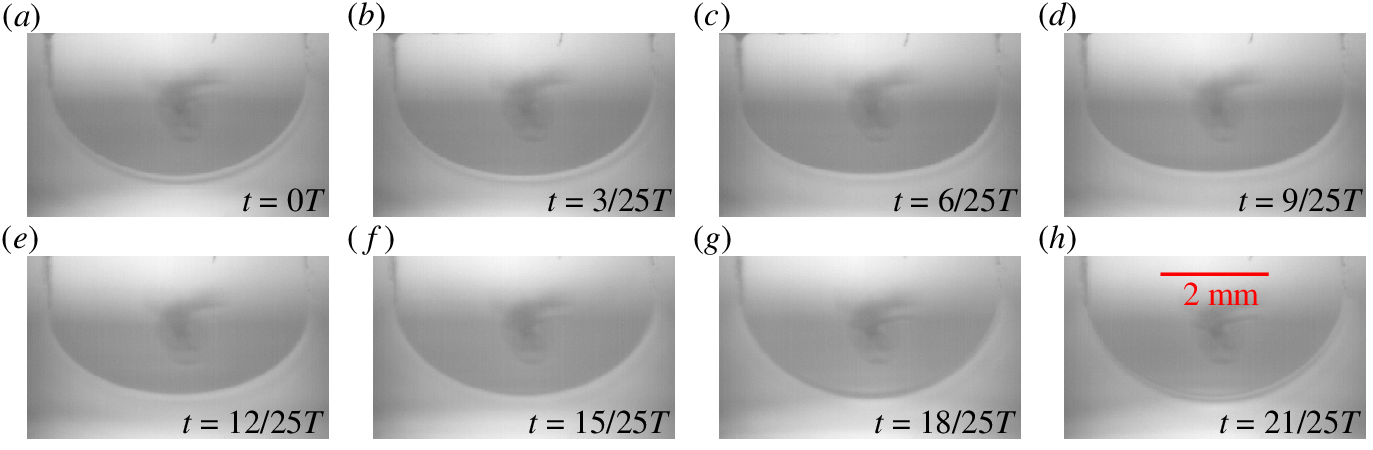}}
\caption{Experimental snapshots of the meniscus for 80$\%$ ethanol solution with $f=30$ Hz and $a=3.9\ \mathrm{m/s^2}$.
The meniscus oscillates synchronously with the external vibration, implying a period $T=1/30$ s.
The contact line is nearly fixed, while the contact angle reaches the boundary of the hysteresis range.
If the acceleration amplitude is further increased to $a=4.0\ \mathrm{m/s^2}$, the contact line will move rapidly and Faraday waves emerge.}
\label{fig:evolution of meniscus}
\end{figure}

The container is filled with the corresponding liquids to a depth of 20 mm and fixed on the vibration generator.
Without external vibration, the static contact angle is extracted.
When the amplitude of the forcing acceleration is lower than the critical threshold, the observed meniscus oscillates synchronously with the external vibration.
Once $a_c$ is reached, Faraday waves emerge, and the meniscus oscillates at a frequency half that of the external vibration.
In this circumstance, the contact line moves rapidly and the contact angle exceeds the hysteresis range.
Therefore, experimental data obtained at acceleration levels just a step (0.1 $\mathrm{m/s^2}$) below the critical threshold $a_c$ are used to extract the hysteresis range.
The temporal evolution of the meniscus profile under specified vibration parameters just before Faraday onset is presented in figure~\ref{fig:evolution of meniscus}.
In this configuration, the contact line remains almost pinned on the lateral walls, while the central region of the meniscus undergoes up-and-down motions.
Near the equilibrium position (figure~\ref{fig:evolution of meniscus}\textit{a},\textit{h}), the meniscus profile closely approximates a circular arc, but at maximum elevation (figure~\ref{fig:evolution of meniscus}\textit{d}), the central region is significantly flattened.

To automatically extract the dynamic contact angle from the high-speed video frame by frame, a customized MATLAB program based on the polynomial fitting approach is developed \citep{kalantarian2011simultaneous,shen2024dynamic}.
For each experimental snapshot, the meniscus profile is extracted by means of the Canny edge detector and then approximated with a sixth-order polynomial curve fitted with 260 pixels.
The contact angle is determined by the derivative of the polynomial fit at the lateral wall positions, followed by averaging the values obtained from both sides of the gap.
We refer to the supplementary material for detailed image processing steps.

\begin{figure}
\centerline{\includegraphics[width=12cm]{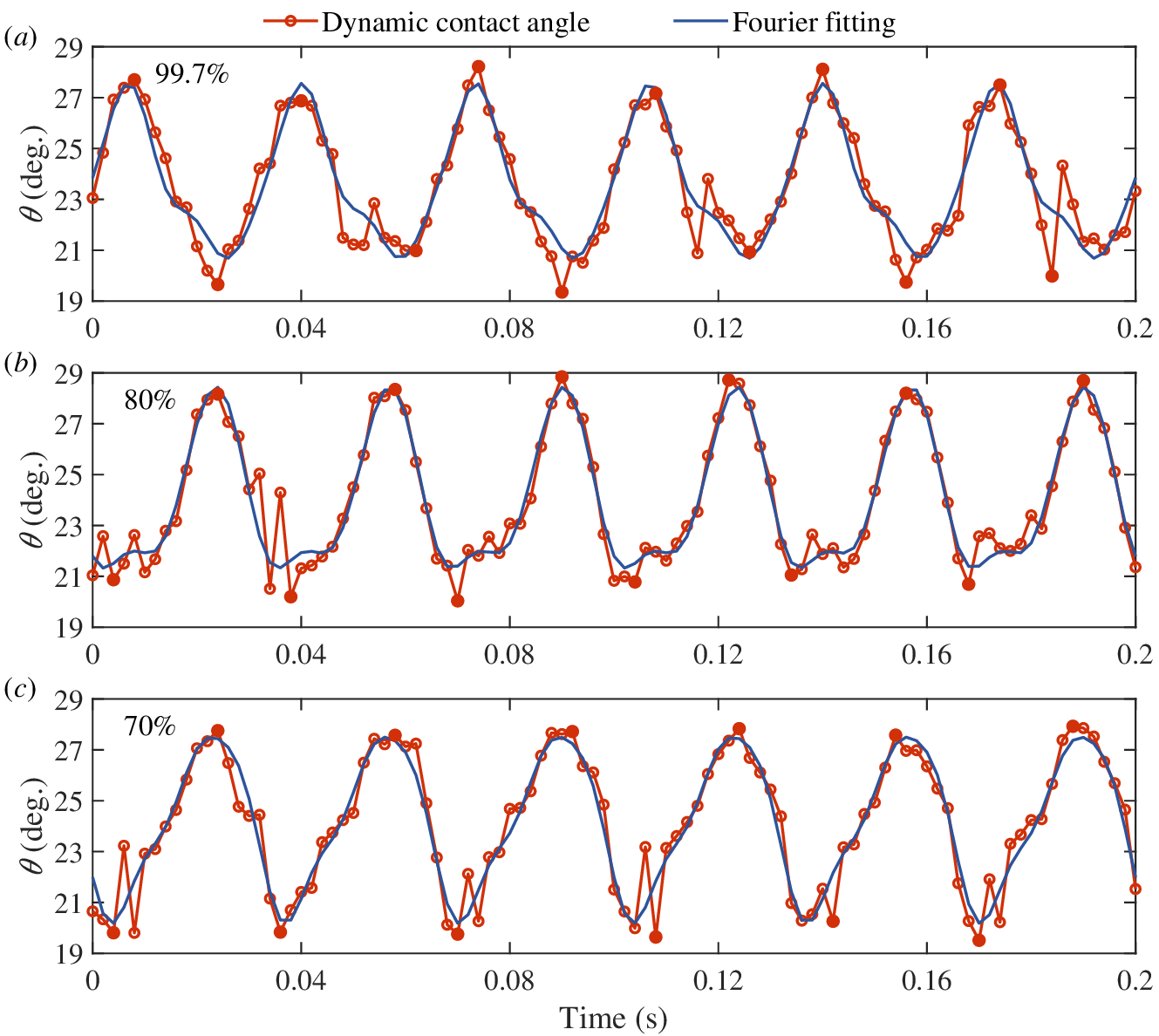}}
\caption{Temporal variation of the contact angle under 30 Hz external vibration for various liquids in the PVC container.
(\textit{a}) Pure ethanol, with $a=3.4\ \mathrm{m/s^2}$.
(\textit{b}) $80\ \text{vol}\%$ ethanol solution, with $a=3.9\ \mathrm{m/s^2}$.
(\textit{c}) $70\ \text{vol}\%$ ethanol solution, with $a=4.1\ \mathrm{m/s^2}$.
Red lines with circles represent for the experimental data in six cycles.
In order to make the trend of $\theta$ more obvious, we perform Fourier fitting to the third-order on these data and eliminate local deviation, which are plotted with blue lines.
Red filled dots identify the maximum and minimum values of $\theta$ in one cycle and correspond to the advancing and receding contact angles.
}
\label{fig:evolution of contact angle}
\end{figure}

Through this technique, we obtain the temporal evolution of the contact angle, which is shown in figure~\ref{fig:evolution of contact angle}.
Fourier-fitted curves are superimposed on the experimental data to suppress high-frequency noise and elucidate the underlying trend.
The contact angle varies periodically between its maximum and minimum values, namely the advancing and receding contact angles.
Far from the limiting values, the contact angle varies rapidly.
As it approaches the hysteresis boundary, the rate of variation decreases significantly.
As a result, these signals exhibit an intermediate shape between sinusoidal and square wave patterns.
This trend is consistent with the expected dynamic wetting behavior observed by \citet{cocciaro1993experimental} and \citet{shen2024dynamic}.

\begin{table}
\begin{center}
\def~{\hphantom{0}}
\begin{tabular}{l@{\hspace{25pt}}c@{\hspace{25pt}}c@{\hspace{25pt}}c@{\hspace{25pt}}c}
Liquid   &$\theta_s\ (\mathrm{deg.})$  & $\theta_a\ (\mathrm{deg.})$ &   $\theta_r\ (\mathrm{deg.})$ & $\rmDelta\ (\mathrm{deg.})$ \\[5pt]
99.7$\%$ pure ethanol & $25.04\pm0.77$ & $27.60\pm0.72$ & $20.11\pm0.88$ & $7.49\pm1.60$\\
80$\%$ ethanol solution & $25.91\pm0.81$ & $28.49\pm0.35$ & $20.60\pm0.57$ & $7.89\pm0.91$ \\
70$\%$ ethanol solution & $26.15\pm1.15$ & $27.73\pm0.20$ & $19.80\pm0.46$ & $7.93\pm0.61$ 
\end{tabular}
\caption{Contact angle parameters of pure ethanol and its solutions in the PVC container.}
\label{tab:contact parameters}
\end{center}
\end{table}

For each kind of liquid, the six maximum values as shown in figure~\ref{fig:evolution of contact angle} are averaged as the advancing contact angle $\theta_a$, while the six minimum values are averaged as the receding contact angle $\theta_r$.
The hysteresis range is then calculated by $\rmDelta=\theta_a-\theta_r$.
The static contact angle $\theta_s$ is extracted from the snapshots that record the static free surface and has been averaged with repeated experiments.
Measurements of these parameters are listed in table~\ref{tab:contact parameters}.
\citet{yang2024ethanol} once studied the wettability property of ethanol--water solutions on hydrophobic highly oriented pyrolytic graphite and found that the contact angle decreases with increasing volume percentage of ethanol, consistent with our measured $\theta_s$.

\begin{figure}
\centerline{\includegraphics[width=13.5cm]{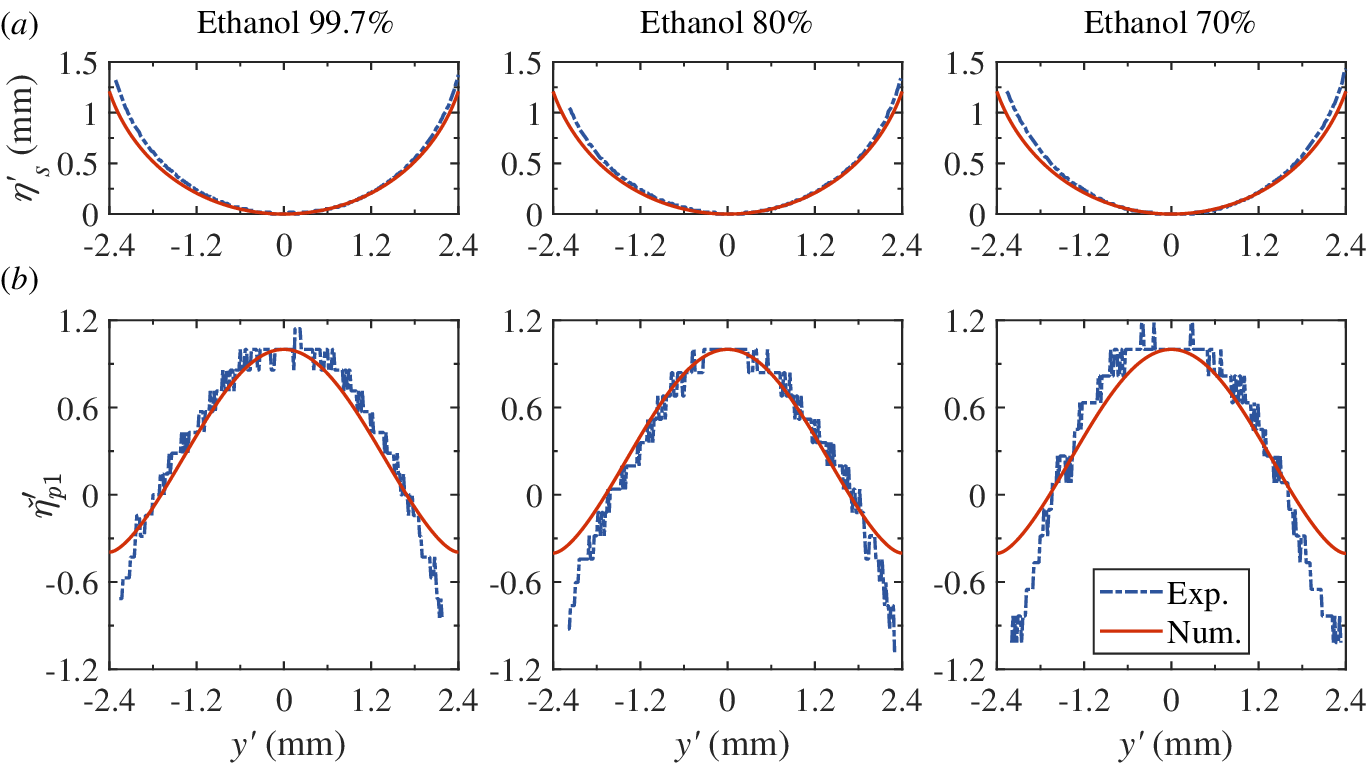}}
\caption{Comparison of meniscus dynamics for three working liquids: experimental observations (chain lines) versus numerical solutions (solid lines).
(\textit{a}) Static meniscus profiles $\eta_s'$.
(\textit{b}) $\epsilon^0$-order results of $\check{\eta}_{p1}'$, normalized by imposing that $\check{\eta}_{p1}' (0 )=1$.}
\label{fig:comparison of meniscus shape}
\end{figure}

Furthermore, meniscus profiles are extracted from experimental snapshots (e.g., figure~\ref{fig:evolution of meniscus}) and compared with numerical results.
The static meniscus is obtained directly from the frames without external vibration.
The maximum oscillatory displacement is then calculated by subtracting this static profile from the extreme meniscus position observed during vibration.
Recalling that the equations governing the static meniscus~\eqref{eq:eta_s} and the $\epsilon^0$-order problem in \S~\ref{sec:Order epsilon^0} are solved numerically, using the experimentally measured contact angle parameters in table~\ref{tab:contact parameters}, we obtain the numerical results of $\eta_s'$ and $\check{\eta}_{p1}'$.
In figure~\ref{fig:comparison of meniscus shape}\textit{a}, theory developed for the static meniscus shows good agreement with experiments, particularly for the lower part.
The heights of the meniscus are nearly identical, which verifies the measured $\theta_s$.
In figure~\ref{fig:comparison of meniscus shape}\textit{b}, results of $\check{\eta}_{p1}'$ are normalized by setting $\check{\eta}_{p1}'(0)=1$.
The shape of the oscillatory meniscus is basically captured by the theory, except for the region near the lateral walls.
At order $\epsilon^0$, the free-edge boundary condition~\eqref{eq:gap O1 contact line condition} is imposed on the free surface, implying a shape always orthogonal to the wall, which is absent in experiments.

\subsection{Faraday onset}
\label{sec:Faraday onset}

\begin{figure}
\centerline{\includegraphics[width=12cm]{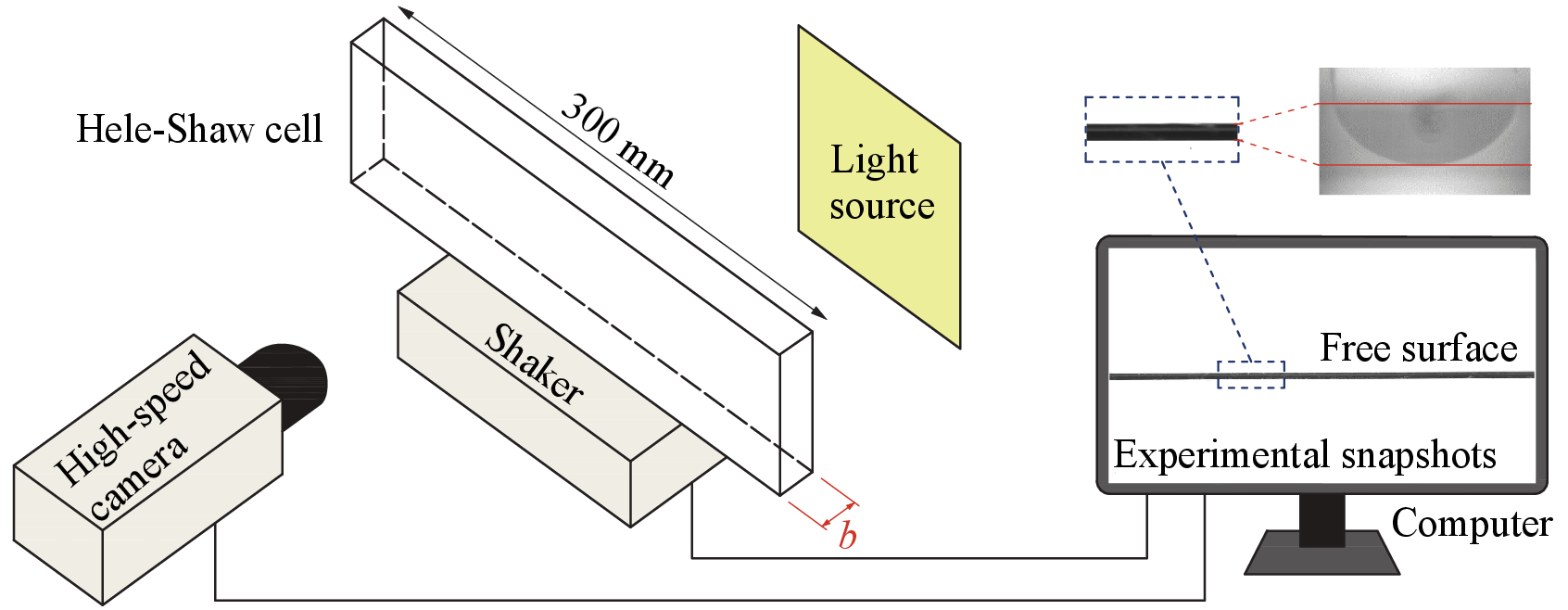}}
\caption{Schematic illustration of the experimental set-up for Faraday instability experiments, with the camera focused on the free surface from a front view of the Hele-Shaw cell.}
\label{fig:set-up_faraday}
\end{figure}

In Faraday instability experiments, traditional rectangular Hele-Shaw cells are utilized (see figure~\ref{fig:set-up_faraday} for the sketch), with a length of 300 mm and a height of 60 mm.
Two different gap sizes are adopted, one of 2 mm and the other of 4.8 mm.
The camera is placed perpendicular to the lateral wall of the cell, hence only the Faraday wave profile is observed.
We concentrate on the critical acceleration amplitude $a_c$ and the critical wavenumber $k_c$ of linear instability of the free surface.
Following the experimental procedure of \citet{douady1988pattern} and \citet{li2019stability}, we gradually increase the external acceleration amplitude $a$ by a step of 0.1 $\mathrm{m/s^2}$ with a fixed driving frequency $f$.
At each level of acceleration amplitude, we wait for 1000 periods of the waves, which is regarded as the growth time for the onset.
When Faraday waves emerge, the acceleration amplitude is recorded as the threshold $a_c$, and the corresponding wavenumber is measured as the critical wavenumber $k_c=2\pi/L_c$, with $L_c$ denoting the critical wavelength of Faraday waves.
Each experiment is repeated three times.

Typically, the threshold measured by gradually raising the driving acceleration corresponds to the lowest point of the Faraday instability tongue (e.g., figure~\ref{fig:Faraday tongue}), which is, in fact, a tricritical point.
Beyond this point, the emerging Faraday waves may be supercritical or subcritical depending on which side of the tongue is considered.
The supercritical branch exhibits a continuous smooth instability with a single reversible threshold, whereas the subcritical branch shows hysteresis, with the onset acceleration exceeding the extinction acceleration.
Such a bifurcation behavior of Faraday waves has been well investigated by \citet{rajchenbach2015faraday}.
% Including the contact angle hysteresis effect might suggest a globally subcritical bifurcation, implying that the measurements depend strongly on whether the acceleration is approached from below or from above.
However, in their derivation the dynamics of contact angle was not considered.
In fact, for our cases, the wetting conditions on the lateral walls from rest to onset and the reverse are different. 
The contact angle hysteresis maters before the contact line is moving.
Once the instability is triggered, the resulting wave amplitude is much higher than that in wide-mouth containers and, most importantly, a liquid film forms on the lateral walls, which was observed both by \citet{li2019stability} and in our experiments.
When such a film is present, there is no longer a sharp contact line between the liquid and the wall.
Even if the acceleration is reduced from a fully developed Faraday wave stage, no clear hysteresis effect exists in wetting dynamics, and the friction force near the lateral wall is significantly diminished, implying that the hysteresis model is not applicable anymore.
Given this, when the acceleration decreases from the instability to the still state the model is apparently different from the one used for the emerging process.
Therefore, we hypothesize that the supercritical branch could vanish for Faraday instability in Hele-Shaw cells.

\begin{figure}
\centerline{\includegraphics[width=11cm]{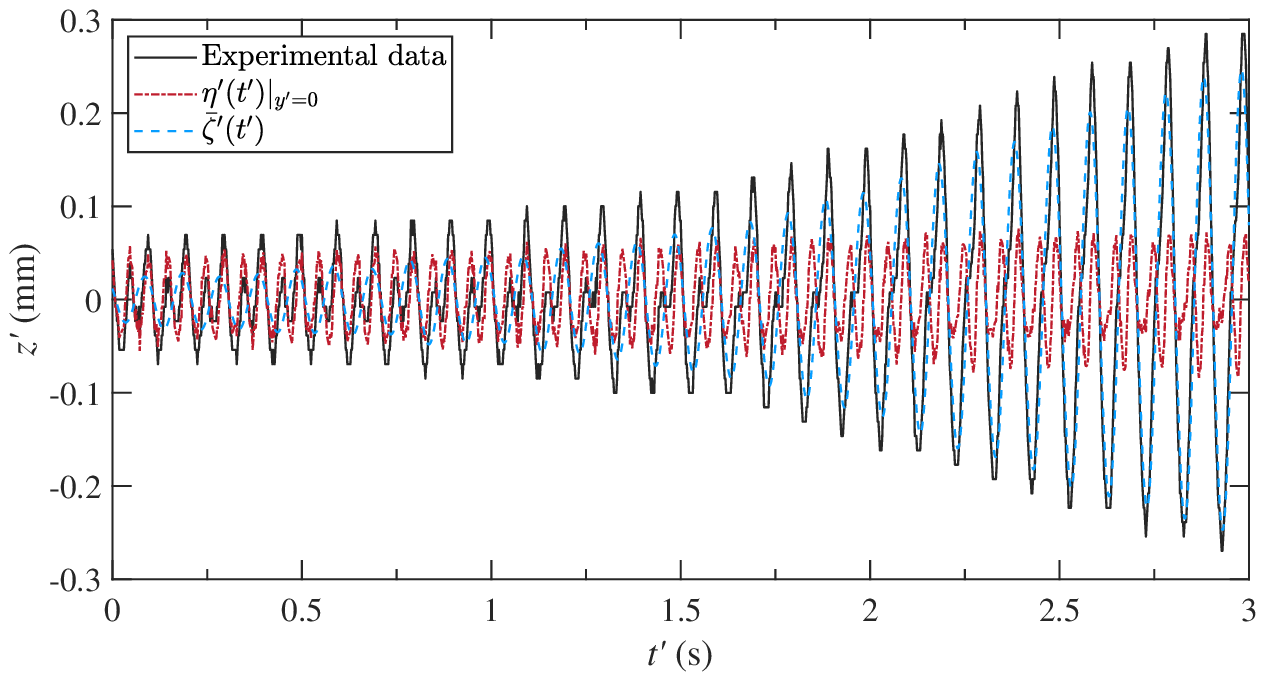}}
\caption{Temporal evolution of the experimentally observed meniscus bottom (black solid line), along with a signal decomposition via fast Fourier transform that extracts the harmonic meniscus wave $\eta'(t')|_{y'=0}$ (red chain line) and the rising subharmonic Faraday wave $\bar{\zeta}'(t')$ (blue dashed line).
Results show the data for $80\%$ ethanol solution in a cell with $b=2\ \mathrm{mm}$, measured at a driving frequency of $f=20\ \mathrm{Hz}$.}
\label{fig:measured B}
\end{figure}

The oscillation amplitude of the meniscus wave $\left | B' \right |$ is measured by the experimental setup used in this section.
From experimental snapshots in Faraday instability experiments (e.g., figure~\ref{fig:set-up_faraday}), the free surface is recorded by a distinct black band with a specific width.
The upper edge of this band corresponds to the contact line between the liquid and the lateral walls, while the lower edge corresponds to the bottom of the meniscus, implying that the width of the black band reflects the height of the meniscus.
The temporal evolution of the bottom of the meniscus is therefore tracked by monitoring the motion of the lower edge of the black band.
Since the linear stability analysis targets the critical condition for Faraday onset, corresponding to the moment when the free surface begins to oscillate subharmonically and the contact line starts to move slightly, $\left | B' \right |$ should be measured when $a=a_c$.
An example of the experimental evolution of the meniscus bottom is plotted in figure~\ref{fig:measured B}.
Before onset, the meniscus oscillates harmonically in response to the forcing (confirmed by figure~\ref{fig:evolution of contact angle}), while no visible wave deformation is observed in the $x'$-direction.
Near the onset, the free surface oscillation amplitude grows gradually as the Faraday instability begins to form, as shown in figure~\ref{fig:measured B}.
This observation interval corresponds to the earliest stage of Faraday instability: the recorded free surface displacement remains below about 0.5 mm, which is much smaller than the amplitude of the fully developed Faraday waves ($\sim$ 10 mm), hence consistent with the regime of our study where the Faraday onset is just beginning to emerge.
The step-like envelope shown in figure~\ref{fig:measured B} is an artifact of measurement uncertainty, mainly resulting from the finite resolution of the imaging system.
According to~\eqref{free surface evolution}, the observed free surface evolution should reflect a superposition of the harmonic meniscus wave and the subharmonic Faraday wave.
After a signal decomposition of the experimental data in figure~\ref{fig:measured B} via fast Fourier transform, the amplitude spectrum exhibits two distinct peaks: one at the harmonic frequency (20 Hz, the driving frequency) and the other at the subharmonic frequency (10 Hz), supporting the decomposition treatment adopted for the three-dimensional free surface and the assumed temporal dependence in the ansatzes for both Faraday and meniscus waves.
The separated signals are also presented in~\ref{fig:measured B}, revealing that the amplitude of the nearly invisible ``Faraday waves'' grows slowly and exponentially while the amplitude of the meniscus wave remains almost constant.
The meniscus wave amplitude in each cycle is extracted and averaged to obtain the measurements of $\left | B' \right |$, with three repeated experiments.
The resulting values of $\left | B' \right |$ for different working liquids and gap sizes, obtained across all experimental driving frequencies, are summarized in table~\ref{tab:|B'| parameters}.

\begin{table}
\begin{center}
\def~{\hphantom{0}}
\begin{tabular}{l@{\hspace{25pt}}c@{\hspace{20pt}}c@{\hspace{20pt}}c@{\hspace{40pt}}c@{\hspace{20pt}}c@{\hspace{20pt}}c}
Driving frequency & \multicolumn{3}{c}{\hspace{-40 pt}$b=2\ \mathrm{mm}$} & \multicolumn{3}{c}{$b=4.8\ \mathrm{mm}$} \\
$f\ (\mathrm{Hz})$ &99.7$\%$ & 80$\%$ &70$\%$ &99.7$\%$ &80$\%$ &70$\%$ \\[5pt]
10 & 0.2011 & 0.1745 & 0.1675 & 0.2927 & 0.2545 & 0.2594 \\
12 & 0.1645 & 0.1425 & 0.1403 & 0.2179 & 0.1885 & 0.1846 \\
14 & 0.1416 & 0.1164 & 0.1132 & 0.1888 & 0.1678 & 0.1573 \\
16 & 0.1155 & 0.0971 & 0.0981 & 0.1665 & 0.1303 & 0.1318 \\
18 & 0.1024 & 0.0833 & 0.0854 & 0.1317 & 0.1160 & 0.1087 \\
20 & 0.0921 & 0.0807 & 0.0795 & 0.1147 & 0.1004 & 0.0935 \\
22 & 0.0808 & 0.0759 & 0.0719 & 0.0968 & 0.0909 & 0.0897 \\
24 & 0.0781 & 0.0705 & 0.0671 & 0.0894 & 0.0865 & 0.0858 \\
26 & 0.0743 & 0.0610 & 0.0634 & 0.0768 & 0.0714 & 0.0704 \\
28 & 0.0715 & 0.0560 & 0.0556 & 0.0682 & 0.0679 & 0.0657 \\
30 & 0.0638 & 0.0512 & 0.0510 & 0.0650 & 0.0633 & 0.0638
\end{tabular}
\caption{Experimental results of the real oscillation amplitude of the meniscus $\left | B' \right |$ measured in mm.
The measurement error is approximately $\pm0.0071$ mm, primarily resulting from the resolution of the imaging system.}
\label{tab:|B'| parameters}
\end{center}
\end{table}

\begin{figure}
\centerline{\includegraphics[width=13cm]{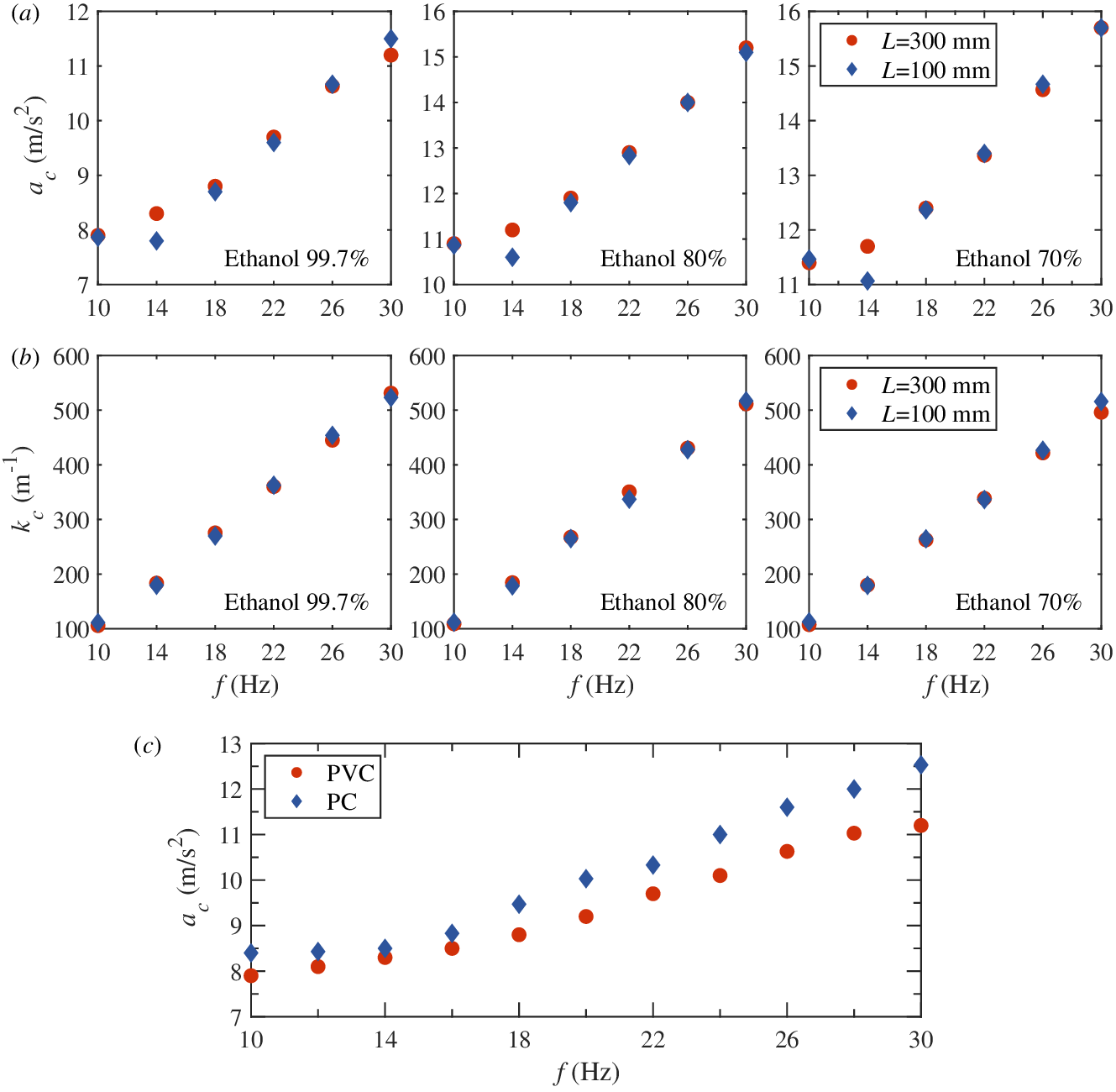}}
\caption{(\textit{a}) Comparison of the experimental results on $a_c$ for different cell length $L$.
Three different ethanol solutions are used, and $b=2$ mm.
(\textit{b}) Comparison of the experimental results on $k_c$  for different $L$.
The working liquids and gap size are consistent with those in (\textit{a}).
(\textit{c}) Comparison of the experimental results on $a_c$ for different materials of the Hele-Shaw cell.
Pure ethanol is used as the working liquid, and $b=2$ mm.}
\label{fig:different length and material}
\end{figure}

In the present theory, the Hele-Shaw cell is assumed to be infinite in length, which results in the periodic condition in $x'$-direction.
However, the presence of two ends in the longitudinal direction may influence the results of linear stability analysis.
\citet{bongarzone2023revised} once employed a thin annulus container to eliminate the effect of two ends.
But they imposed that the azimuthal wavenumber is always an integer.
Hence, these experiments are limited for identifying the change of $a_c$ and $k_c$ induced by the ends of the cell.
Here, we employ a shorter Hele-Shaw cell, which has a length of 100 mm.
In this case, the container is not so long compared with the wavelength, and the influence of the ends is more significant.
As shown in figure~\ref{fig:different length and material}\textit{a}, for different cell lengths, the critical acceleration $a_c$ is generally consistent, except for the case of $f=14$ Hz.
At this driving frequency, the experimental wavelength $L_c$ is not small relative to the cell length $L$, and $L$ is closest to an integer multiple of $L_c$ ($L\approx 2.85 L_c$).
This unique length-scale relationship, absent at other frequencies, may explain the observed discrepancy between different $L$ at $f=14$ Hz.
When the cell length is increased to 300 mm, the impact of the ends can be basically ignored.
A more detailed examination of this phenomenon lies outside the scope of this work.
When it comes to the critical wavenumber, from figure~\ref{fig:different length and material}\textit{b}, we can conclude that the ends of the cell do not affect the dispersion relation at all.
Based on the comparisons of $a_c$ and $k_c$ between two different cell lengths, the use of containers with a length of 300 mm guarantees that the Hele-Shaw cell is sufficiently long to make the effect of two ends negligible when modeling the Faraday instability problem.

As we have emphasized before, contact angle hysteresis damping plays an important role in the threshold of Faraday instability and is related to the static contact angle and hysteresis range.
These parameters depend on both the liquid \citep{yang2024ethanol} and the solid substrate \citep{shen2024dynamic}.
If we change the material of the Hele-Shaw cell, the contact angle parameters will naturally change, leading to the variation of $a_c$.
This may give intuitive evidence of the impact of dynamic wetting on Faraday waves.
Therefore, we conduct experiments in Hele-Shaw cells made of different materials, one of PVC and the other of polycarbonate (PC).
As shown in figure~\ref{fig:different length and material}\textit{c}, with only the material changing, the values of $a_c$ are different.
This discrepancy cannot be attributed to viscosity or gap-averaged damping, and indicates the significance of contact angle on the Faraday instability problem in Hele-Shaw cells.
The comparison also reflects the limitation of the widely used Hamraoui's model~\eqref{Hamraoui model} by \citet{li2019stability}, \citet{bongarzone2023revised}, and \citet{li2024stability}, in which the friction coefficient $\beta$ is only determined by the liquid \citep{hamraoui2000can}.
This dynamic contact angle model cannot explain the discrepancy in figure~\ref{fig:different length and material}\textit{c}.
In addition, noting that $\beta=0.04\ \mathrm{Pa\ s}$ for pure ethanol is obtained using glass capillaries as elucidated by \citet{hamraoui2000can}, it is not rigorous to directly apply the experimental measurements of $\beta$ in glass containers to the contact angle dynamics in PVC containers.
These wetting parameters must be experimentally measured using containers made of identical materials to those employed in Faraday instability experiments.

\section{Results and discussion}
\label{sec:Results and discussion}

\subsection{Stability analysis}
\label{sec:Stability analysis}

In \S~\ref{sec:Amplitude equation for Faraday waves}, an amplitude equation for Faraday instability is derived, which contains the damping of the fluid viscosity and the confinement of two lateral walls.
Meanwhile, in \S~\ref{sec:Contact angle hysteresis damping}, another amplitude equation is derived that contains only the damping of contact angle hysteresis effects.
All of these damping sources should be taken into account in stability analysis.
Because the damping constitutes an intrinsic property of the system that remains independent of amplitude, it is feasible to incorporate the damping term in~\eqref{eq:non-dimen gap amplitude equation} directly into~\eqref{amplitude equation for faraday}, which results in the combined amplitude equation:
\begin{equation}
\frac{\mathrm{d} A}{\mathrm{d} T} +\frac{1}{2} \left [ \mathrm{i} \left ( \hat{\delta}_{1,r}+\mathrm{i} \hat{\delta}_{1,i} \right )+\mathrm{i}\omega_1^2+2\hat{\delta} +2\chi  \right ] A+\frac{\mathrm{i}\hat{a}}{4g\left ( 1+l_c^2k^2 \right ) }A^*=0.
\label{eq:combined amplitude equation}
\end{equation}

We employ Lyapunov's first method to examine the stability of the zero solution of the autonomous dynamic system~\eqref{eq:combined amplitude equation}.
This trivial solution represents the rest state of the free surface.
When the forcing acceleration exceeds a critical threshold, the rest state cannot remain stable, giving rise to Faraday waves.
Around the zero solution, we introduce the perturbation $\Lambda_i$, with $i=1,2$ corresponding to~\eqref{eq:combined amplitude equation} and its conjugate equation.
The perturbation satisfies
\begin{subequations}
\begin{equation}
\frac{\mathrm{d} \Lambda _1}{\mathrm{d} T} =-\frac{1}{2} \left ( 2\hat{\delta} +2\chi - \hat{\delta}_{1,i} + \mathrm{i} \hat{\delta}_{1,r} +\mathrm{i}\omega_1^2 \right ) \Lambda _1-\frac{\mathrm{i}\hat{a}}{4g\left ( 1+l_c^2k^2 \right ) }\Lambda _2,
\label{eq:xi1}
\end{equation}
\begin{equation}
\frac{\mathrm{d} \Lambda _2}{\mathrm{d} T} =-\frac{1}{2} \left ( 2\hat{\delta} +2\chi - \hat{\delta}_{1,i}-\mathrm{i} \hat{\delta}_{1,r} -\mathrm{i}\omega_1^2 \right ) \Lambda _2+\frac{\mathrm{i}\hat{a}}{4g\left ( 1+l_c^2k^2 \right ) }\Lambda _1.
\label{eq:xi2}
\end{equation}
\end{subequations}
Assuming $\Lambda _i=\Lambda _{i,0}e^{\lambda T}$, an eigenvalue equation is obtained that
\begin{equation}
\lambda ^2+ \left ( 2\hat{\delta }+ 2\chi -\hat{\delta }_{1,i}   \right ) \lambda +\frac{1}{4} \left [ \left ( 2\hat{\delta } +2\chi -\hat{\delta }_{1,i} \right )^2+\left (\hat{\delta }_{1,r}+\omega _1^2 \right )^2   \right ]-\frac{\hat{a} ^2}{16g^2\left (1+l_c^2k^2  \right )^2 }=0.
\label{eq:eigenvalue equation}
\end{equation}
The eigenvalue is solved as
\begin{equation}
\lambda _{\pm}=\frac{1}{2}\left [ -2\hat{\delta }-2\chi +\hat{\delta }_{1,i}\pm \sqrt{\frac{\hat{a} ^2}{4g^2\left (1+l_c^2k^2  \right )^2 }-\left ( \hat{\delta }_{1,r} +\omega_1^2\right ) ^2}  \right ].
\label{eq:lambda}
\end{equation}
The term $\hat{\delta }_{1,i}$ is relatively small compared with other terms in~\eqref{eq:lambda}.
Hence, the solution $\lambda_-$ is certainly negative.
In order to make the rest state asymptotically stable, $\lambda_+$ must be negative, which leads to a condition for $\hat{a}$:
\begin{equation}
\left | \hat{a} \right |<  2g \left (1+l_c^2k^2  \right )\sqrt{\left ( 2\hat{\delta } +2\chi -\hat{\delta }_{1,i} \right )^2 +\left ( \hat{\delta }_{1,r} +\omega_1^2\right ) ^2}.
\label{eq:a_hat}
\end{equation}
For the periodic vertical vibration, different signs of $\hat{a}$ represent different directions of the acceleration.
By multiplying the parameter $\epsilon$ to~\eqref{eq:a_hat}, and combining with the definition~\eqref{parameter expanding of Faraday} and~\eqref{eq:non-dimen gap amplitude equation}, for a specific value of $k$, an analytical expression for the homologous acceleration amplitude $a_k$ is obtained that
\begin{equation}
a_k= 2g \left (1+l_c^2k^2  \right )\sqrt{\left ( 2\delta _{St}^2k^2b^2 +\frac{4\mathrm{i}\sin\theta_s\rmDelta \sigma \kappa}{\pi \rho \Omega | \check{\eta}_{p1}' |_{y'=b/2} \left | B' \right | } -\delta_{1,i} \right )^2 +\left ( \delta_{1,r} +\omega^2-1\right ) ^2}.
\label{eq:ak}
\end{equation}

\begin{figure}
\centerline{\includegraphics[width=11cm]{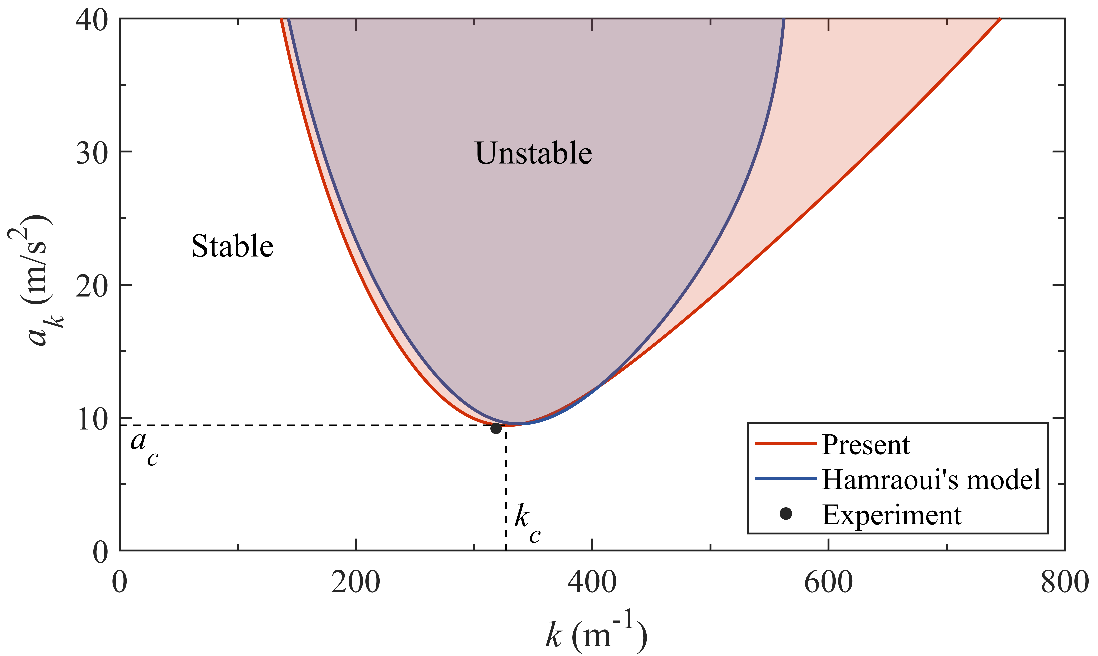}}
\caption{First subharmonic Faraday tongue for pure ethanol in a Hele-Shaw cell of $b=2$ mm, with $f=20$ Hz.
The red unstable region is obtained via~\eqref{eq:ak} while blue region is computed by reproducing the Floquet analysis developed by \citet{bongarzone2023revised} with the employment of Hamraoui's model ($\beta=0.04$ Pa s).
Experimental data for this configuration is identified by the black dot.
The lowest points on the boundary curves determine the critical parameters $a_c$ and $k_c$ for Faraday instability.}
\label{fig:Faraday tongue}
\end{figure}

Equation~\eqref{eq:ak} reveals that $a_k$ depends on the fluid properties, contact angle characteristics, and external parametric excitation.
For a given set of these parameters, $a_k$ varies solely with $k$.
The value of $a_k$ corresponding to the critical wavenumber $k_c$ is the critical acceleration amplitude $a_c$.
Using the method given by \citet{li2024stability}, we traverse the values of $k$ in a range of 0 $\text{m}^{-1}$ -- 1000 $\text{m}^{-1}$, after which $a_k$ is computed via~\eqref{eq:ak}, the minimum of $a_k$ is then identified as $a_c$, and its corresponding wavenumber is regarded as $k_c$.
Following this procedure, the first subharmonic Faraday instability boundary is obtained, as shown in figure~\ref{fig:Faraday tongue}.
Its shape is qualitatively similar to the tongue-like instability region reported by \citet{kumar1994parametric}.
For comparison, we include theoretical results obtained by reproducing the Floquet analysis of \citet{bongarzone2023revised}, which adopted Hamraoui's dynamic contact angle model and neglected viscous dissipation.
In the present experimental configuration, the threshold $a_c$ obtained by both theories is in good agreement with the experimental one.
However, the critical wavenumber $k_c$ obtained by \citet{bongarzone2023revised} shifts to a higher value compared to the experimental one, while the present theory agrees better with the experiment.
This frequency detuning is partially compensated by the current theoretical framework.
A detailed discussion on the dispersion relation of Faraday waves will be given in the following section.

\begin{figure}
\centerline{\includegraphics[width=13.5cm]{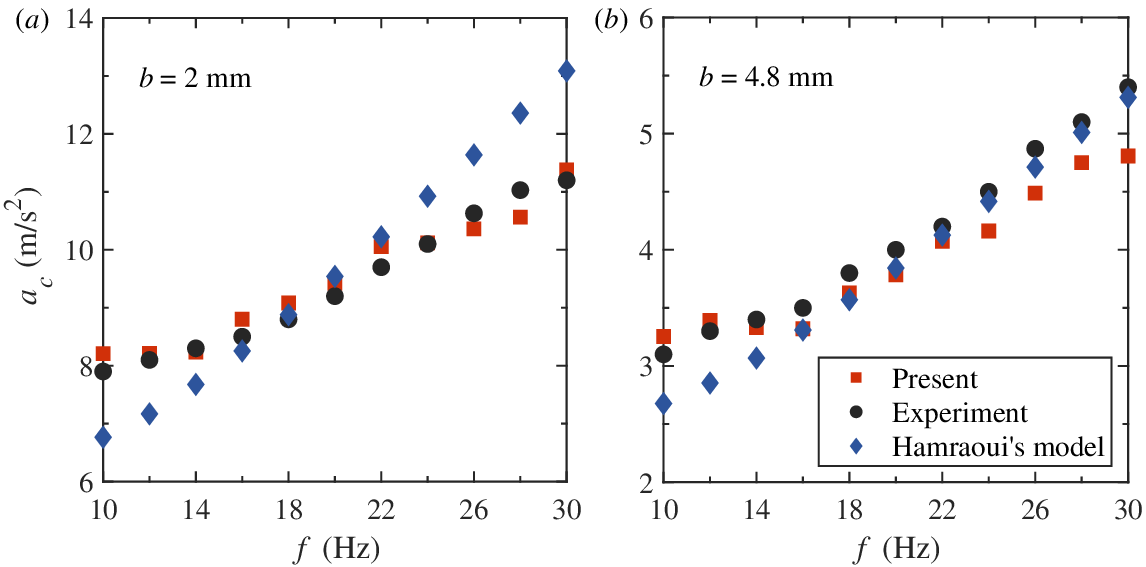}}
\caption{Critical acceleration amplitude $a_c$ for the onset of Faraday waves versus experimental driving frequency $f$.
Comparison between theoretical results (squares and diamonds represent results calculated by the present theory and Hamraoui's model used by \citet{bongarzone2023revised}, respectively) and experimental measurements (circles).
The working liquid is pure ethanol, with $\beta=0.04$ Pa s.
The gap size is $b=2$ mm for (\textit{a}), and $b=4.8$ mm for (\textit{b}).}
\label{fig:critical acceleration}
\end{figure}

\begin{figure}
\centerline{\includegraphics[width=7cm]{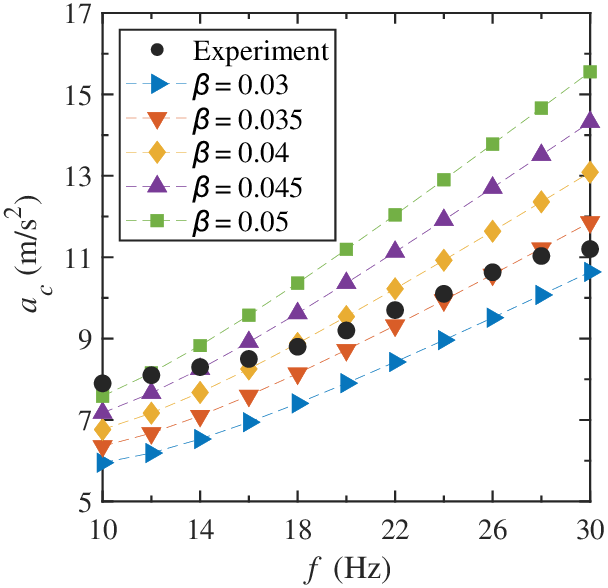}}
\caption{Critical acceleration amplitude $a_c$ obtained using Hamraoui's model and experimental measurements.
Different values of $\beta$ reported in the figure are used to calculate $a_c$.
The working liquid is pure ethanol, while $b=2$ mm.}
\label{fig:change beta}
\end{figure}

Now, let us examine the theoretical results of $a_c$ with experiments in a wide range of driving frequency, which are summarized in figure~\ref{fig:critical acceleration}.
\citet{li2019stability} once performed Faraday instability experiments in Hele-Shaw cells in a frequency range of 14--22 Hz.
We have extended this range to 10--30 Hz to highlight the trend of $a_c$.
Theoretically, \citet{bongarzone2023revised} employed Stokes boundary layer theory when studying the linear stability problem in Hele-Shaw cells, coupled with Hamraoui's model~\eqref{Hamraoui model} to evaluate the damping of the dynamic contact angle.
Their theory has currently provided the most accurate estimation of $a_c$.
Figure~\ref{fig:critical acceleration} compares the experimentally measured acceleration thresholds for Faraday onset with theoretical results obtained from the present theory~\eqref{eq:ak} and from the reproduced analysis of \citet{bongarzone2023revised}.
With the utilization of Hamraoui's model, their theory agrees well with experiments, particularly for the case of $b=4.8$ mm.
This better agreement for the larger gap size is likely due to the reduced influence of meniscus dynamics; specifically, the surface tension and the pinned contact line effect in the gap direction weaken as the gap becomes wider.
However, a notable discrepancy in the trend persists and is more evident for a smaller gap size ($b=2$ mm), where the two aforementioned effects play a more prominent role.
Hamraoui's model predicts a steeper variation of $a_c$ with $f$, resulting in significantly overestimated theoretical values at high driving frequencies.
\citet{bongarzone2023revised} have validated that the increasing trend of $a_c$ with $f$ is mainly controlled by the friction coefficient $\beta$, a conclusion also verified by \citet{li2024stability}.
Equation~\eqref{Hamraoui boundary} implicitly reveals that $\beta$ serves as a linear coefficient for contact angle damping, implying that the theoretical $a_c$ is sensitive to $\beta$.
As validated in figure~\ref{fig:different length and material}\textit{c}, the value of $\beta$ for pure ethanol in PVC containers must be different from the one measured by \citet{hamraoui2000can}.
Figure~\ref{fig:change beta} demonstrates that even modest variations of $\beta$ ($0.03 \le \beta \le 0.05$) in the theory of \citet{bongarzone2023revised} produce substantial changes in the calculated $a_c$, which obviously deviate from the experimental results.
Consequently, Hamraoui's model lacks physical justification for the Faraday instability problem in Hele-Shaw cells, with its successful application critically depending on values of $\beta$.

\begin{figure}
\centerline{\includegraphics[width=13.5cm]{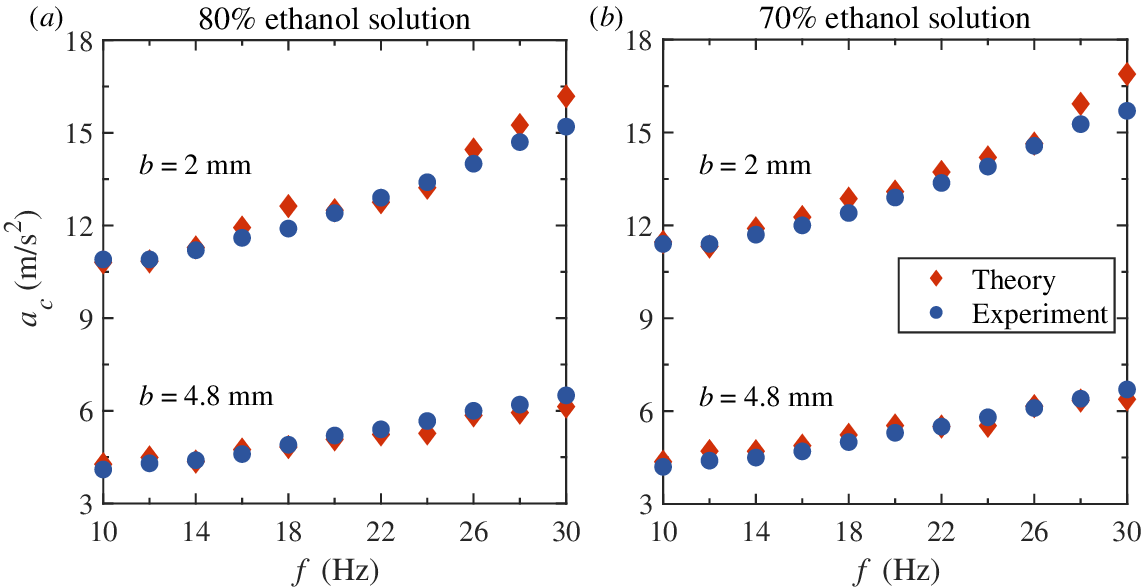}}
\caption{Critical acceleration amplitude $a_c$ for the onset of Faraday waves versus experimental driving frequency $f$.
Comparison between the present theoretical results (diamonds) and experimental measurements (circles).
The working liquid is 80\% ethanol solution for (\textit{a}) and 70\% ethanol solution for (\textit{b}).
Upper regions correspond to the case of $b=2$ mm, while lower regions correspond to the case of $b=4.8$ mm.}
\label{fig:critical acceleration for solution}
\end{figure}

In contrast, our model adequately takes into account unsteady Stokes flow dynamics, pinned contact line constraints, and full viscous dissipation effects.
The contact angle hysteresis damping is obtained by resolving the gap flow independently.
According to~\eqref{eq:ak}, the trend of $a_c$ versus $f$ is determined by the static contact angle $\theta_s$, the hysteresis range $\rmDelta$, the shape of the meniscus reflected in $\kappa$ and $| \check{\eta}_{p1}' |_{y'=b/2}$, and the oscillation amplitude of the meniscus $|B'|$.
Since we have experimentally measured these parameters and solved for $\kappa$ and $| \check{\eta}_{p1}' |_{y'=b/2}$ numerically, the present theory better meets the Faraday instability problem.
For $b=2$ mm (figure~\ref{fig:critical acceleration}\textit{a}), the trend of $a_c$ shows an improved agreement with the experiments, indicating that compared to Hamraoui's model, the contact angle hysteresis formulation more accurately captures meniscus dynamics.
In this configuration, the capillary effects are more prominent.
For $b=4.8$ mm (figure~\ref{fig:critical acceleration}\textit{b}), the proposed theory exhibits an agreement similar to that of Hamraoui's model, with both characterizing the Faraday instability well.
Finally, we compare the theoretical results of $a_c$ with experimental measurements in varying ethanol concentrations and gap sizes.
The good agreement shown in figure~\ref{fig:critical acceleration for solution} verifies the applicability of the present model.

\subsection{Dispersion relation}
\label{sec:Dispersion relation}

\begin{figure}
\centerline{\includegraphics[width=13.5cm]{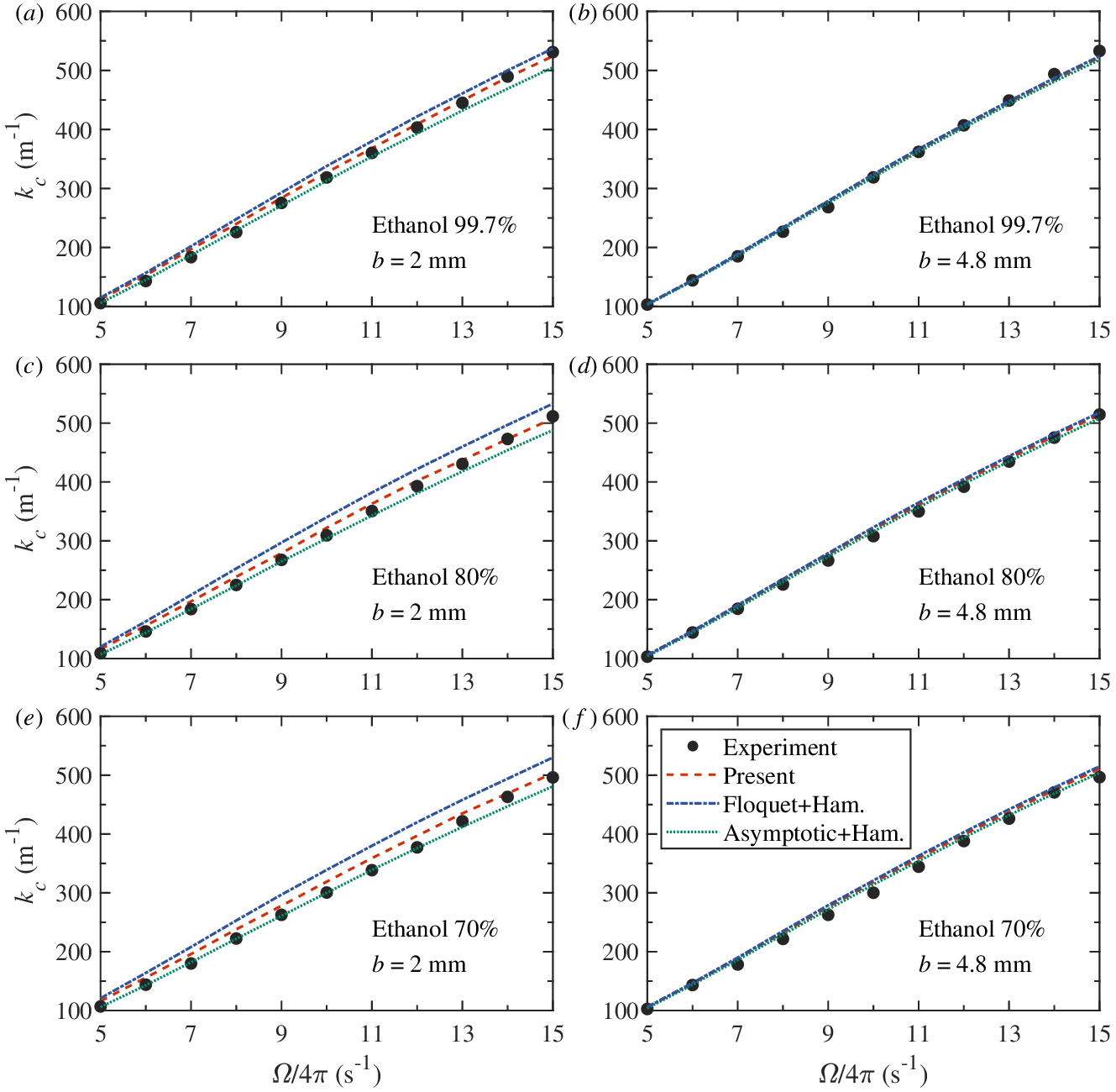}}
\caption{Dispersion relation across different working liquids and gap sizes $b$, which are reported in the figure.
Comparison between experimental measurements (black dots) and theoretical results (red dashed lines: present theory~\eqref{eq:ak}; blue chain lines: Floquet analysis of the full gap-averaged model; green dotted lines: asymptotic expansion of the full gap-averaged model, i.e.~\eqref{eq:ak_ham}; with $\beta=0.04$ (Pa s) for pure ethanol, $\beta=0.0538$ for 80$\%$ ethanol solution, and $\beta=0.0597$ for 70$\%$ ethanol solution).
The $x$-axis is the response frequency of Faraday waves in subharmonic mode, which is half of the driving frequency $f=\Omega/2\pi$.}
\label{fig:critical wavenumber}
\end{figure}

The dispersion relation serves as a fundamental characteristic of Faraday waves, providing crucial insights into their wavelength selection when the instability develops.
\citet{rajchenbach2015faraday} investigated the dispersion relation of Faraday waves and searched for a weakly nonlinear correction to the classical one for gravity-capillary waves with the incorporation of forcing, damping, and nonlinear effects, thereby establishing an effective dispersion relation.
Many studies (e.g., \citet{rajchenbach2011new}, \citet{li2018faraday}, and \citet{li2019stability}) have attempted to apply such modified relations to Faraday waves in Hele-Shaw cells.
Although these formulations align with the experimental regime where Faraday waves are fully developed and nonlinear effects certainly influence wave selection, including the nonlinear correction of the form $K\left | A \right |^2$ does not improve the predictions, reflecting limitations in the Hele-Shaw configurations where liquid films are present \citep{li2019stability}.
In practice, the classical relation $\left (\Omega/2 \right )^2=gk_c+\sigma k_c^3/\rho$ remains accurate, as validated by \citet{li2024stability}.
We have also confirmed that the experimental measurements of $k_c$ rigorously obey this original dispersion relation throughout the driving frequency range.
However, the employment of the oscillatory Stoke boundary layer leads to a frequency detuning, as evidenced by the apparent discrepancy between the Floquet analysis and the experiment in figure~\ref{fig:critical wavenumber} for the case of $b=2$ mm.
\citet{bongarzone2023revised} also reported a discrepancy between response frequency and driving frequency and found that the Faraday instability tongue shifts toward higher critical wavenumbers (see Figure 4 in \citet{bongarzone2023revised}).
Through asymptotic analysis, they demonstrated that the observed detuning stems from the imaginary component of the gap-averaged damping $\delta_{1,i}$, indicating that it is a physically justified and expected phenomenon.
Their experiments in a thin annular container confirmed that this frequency shift is necessary for accurately predicting the locations of Faraday tongues.
In rectangular Hele-Shaw cells, however, neither the experiments of \citet{li2019stability} nor ours exhibit such a phenomenon.
This suggests that additional physical mechanisms are required to address the detuning.

\begin{figure}
\centerline{\includegraphics[width=13.5cm]{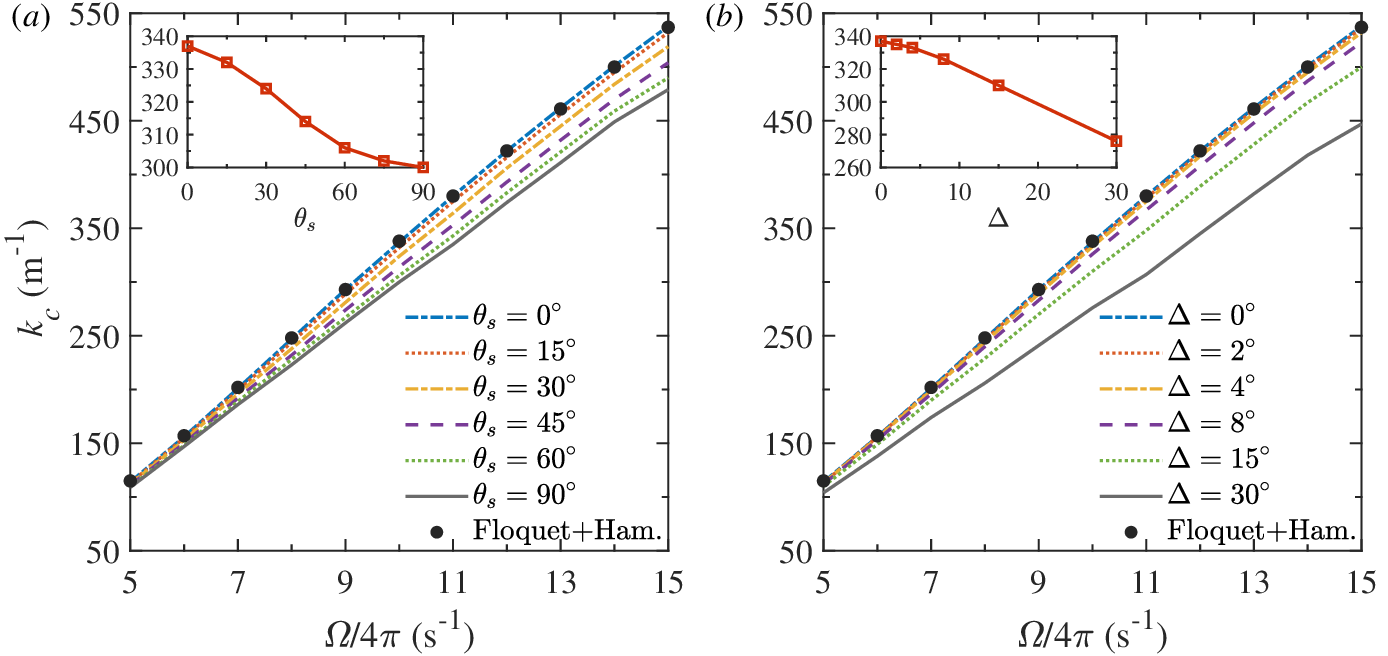}}
\caption{Theoretical dispersion relation for different contact angle parameters.
(\textit{a}) The static contact angle varies from $0^\circ$ to $90^\circ$, with a hysteresis range $\rmDelta=7.49^\circ$.
The inset in (\textit{a}) shows the critical wavenumber $k_c$ versus $\theta_s$ at a response frequency $\Omega/4\pi=10$ Hz.
(\textit{b}) The hysteresis range varies from $0^\circ$ to $30^\circ$, with a static contact angle $\theta_s=25.04^\circ$.
The inset in (\textit{b}) shows $k_c$ versus $\rmDelta$ at a response frequency $\Omega/4\pi=10$ Hz.
Fluid parameters of pure ethanol in table~\ref{tab:fluid parameters} are employed, with $b=2$ mm.
For comparison, the dispersion relation in figure~\ref{fig:critical wavenumber}\textit{a} obtained from Floquet analysis of the full gap-averaged model} is plotted with solid dots.
\label{fig:dispersion relation via contact parameters}
\end{figure}

The Faraday tongue in figure~\ref{fig:Faraday tongue} indicates that the present theory partially compensates for the frequency detuning observed in the results of \citet{bongarzone2023revised}, with both contact angle hysteresis effect and viscous dissipation retained and Floquet analysis replaced by an asymptotic expansion.
In view of these substantial changes in the theoretical framework, we have continued the derivation of \S~\ref{sec:Incorporating meniscus effects into gap-averaged equations} to obtain asymptotic results based on Hamraoui's model (see Appendix~\ref{Linear stability analysis based on Hamraoui's model}) and performed a Floquet analysis of the ``full gap-averaged model'', i.e., contact angle dynamics described by Hamraoui's model and viscous dissipation retained (details in the supplementary material).
Figure~\ref{fig:critical wavenumber} presents the dispersion relation obtained from the present theory~\eqref{eq:ak} and from either Floquet analysis or asymptotic expansion applied to the full gap-averaged model.
The Floquet results (blue chain curves) essentially coincide with the reproduced results of \citet{bongarzone2023revised}.
Overall, the asymptotic expansion based on either contact angle models agrees better with the experiments than does the Floquet analysis.
Comparing the results obtained with different contact angle models, we observe that the contact angle hysteresis model gives a slightly smaller detuning compensation, resulting from the updated contact angle damping term in~\eqref{eq:ak}, yet both formulations achieve nearly the same level of agreement with experiments.
Regarding the influence of gap size, figure~\ref{fig:critical wavenumber} shows that all theoretical results match the experiments better for $b=4.8$ mm than their counterparts, where the impact of meniscus dynamics on the dispersion relation is significantly reduced.
To identify the origin of the detuning compensation observed in the asymptotic results, 
after a straightforward mathematical procedure from~\eqref{eq:ak}, we obtain
\begin{equation}
\left ( \omega^2-1+\delta_{1,r} \right ) = \pm \sqrt{ \frac{a^2}{4g^2 \left (1+l_c^2k^2  \right )^2} -\left ( 2\delta _{St}^2k^2b^2 +\frac{4\mathrm{i}\sin\theta_s\rmDelta \sigma \kappa}{\pi \rho \Omega | \check{\eta}_{p1}' |_{y'=b/2} \left | B' \right | } -\delta_{1,i} \right )^2}.
\label{eq:detuning}
\end{equation}
On the one hand, this equation reveals how Stokes boundary layer theory leads to significant frequency detuning through the term $\delta_{1,r}$.
On the other hand, the right-hand side of~\eqref{eq:detuning} clarifies the mechanism of the detuning compensation: the introduced damping can effectively reduce that detuning.
Indeed, in the present regime, characterized by the Hele-Shaw limit $k^2b^2\ll 1$ and low viscosity, the contribution of viscous dissipation $2\delta _{St}^2k^2b^2$ is negligible.
Regarding the influence of contact angle hysteresis damping, we present a parametric analysis of the static contact angle and the hysteresis range.
Figure~\ref{fig:dispersion relation via contact parameters} shows that as $\theta_s$ or $\rmDelta$ increases, the resulting $k_c$ shifts to noticeably lower values.
The limiting cases $\theta_s=0^\circ$ and $\rmDelta=0^\circ$ correspond to the absence of contact angle hysteresis damping, nearly consistent with the Floquet results for the full gap-averaged model.
The insets of the figure indicate that the decreasing trends of $k_c$ follow roughly the functional dependencies in~\eqref{eq:detuning}, namely proportional to $\sin \theta_s$ and $\rmDelta$.
According to~\eqref{eq:detuning}, raising $\theta_s$ or $\rmDelta$ increases the overall damping, which shifts the dispersion relation towards lower $k$, thus partially reducing the detuning.
% It should be clarified that although the contact angle hysteresis model accounts for the pinned contact line effect, the resulting detuning compensation arises solely from the added damping term, rather than from a direct modification to the frequency that would appear on the left-hand side of~\eqref{eq:detuning}.
% This is because the leading order eigen-solution assumes a free-edge contact line condition~\eqref{eq:gap O1 contact line condition}, which itself introduces no frequency shift.}

Although the Floquet analysis also includes both the linear viscous damping and the contact angle damping based on Hamraoui's model, the corresponding change in $k_c$ remains negligible (see figures~\ref{fig:critical wavenumber} and~\ref{fig:dispersion relation via contact parameters}), indicating that it introduces almost no detuning compensation.
Therefore, this compensation effect is only pronounced in the asymptotic approach.
This difference likely stems from the small-parameter assumption adopted in the asymptotic procedure: with the rescaling treatments listed in~\eqref{parameter expanding of Faraday}, the influence of $\hat{\delta}_1$ appears at a higher order $\textit{O}(\epsilon^2)$ and therefore does not enter the amplitude equation.
This ordering is justified because $\epsilon \hat{\delta}_1=\left ( 1-\mathrm{i} \right )\delta _1 \tanh\left [ \left ( 1+\mathrm{i} \right )/2\delta _1  \right ]$ has a magnitude of order $\delta_1$, which is sufficiently small compared to $\textit{O} (1)$.
% According to~\eqref{aqq:eq:gap dimensionless normal stress condition}, these small effects are primarily coupled with the contact angle dynamics and viscous dissipation.
These components neglected in the amplitude equation may contain imaginary part which will in turn amplify the frequency shift, a similar consequence shown on the left-hand side of~\eqref{eq:detuning}.
Hence, the present asymptotic expansion experiences a lower detuning feature than Floquet method does.
Please note that, as for the proposed gap-resolved model, the asymptotic expansion cannot be circumvented.
% by rescaling the physical parameters and relegating weaker contributions to higher orders, effectively amplifies the frequency modification originating from contact angle dynamics and viscous dissipation.}

It should be noted that a slight detuning remains evident in figure~\ref{fig:critical wavenumber}.
This discrepancy is probably attributed to the fact that the proposed theory only captures the initial stage of Faraday instability.
The theory aims to characterize the critical condition for linear Faraday instability, corresponding to the moment at which the free surface begins to rise rapidly from a still state.
According to~\eqref{eq:gap contact angle}, only the hysteresis component is considered when modeling the contact angle dynamics, revealing a pinned contact line.
However, because the wave pattern arises quickly once the Faraday instability emerges, the experimental data of $k_c$ are necessarily measured after the onset, during which Faraday waves are already fully developed.
In this stage, experimental observations show that the liquid fully wets the lateral walls and a liquid film is formed as the free surface fluctuates, a phenomenon also reported by \citet{li2019stability}.
This change in the wettability condition renders the contact angle hysteresis model no longer applicable.
To address the detuning observed in figure~\ref{fig:critical wavenumber}, alternative mathematical models that better align with the experimental measurement conditions may be effective.
Additionally, nonlinear effects associated with large amplitude Faraday waves may also contribute to the observed discrepancy.

\subsection{Damping in Hele-Shaw system}
\label{sec:Damping in Hele-Shaw system}

\begin{figure}
\centerline{\includegraphics[width=12.5cm]{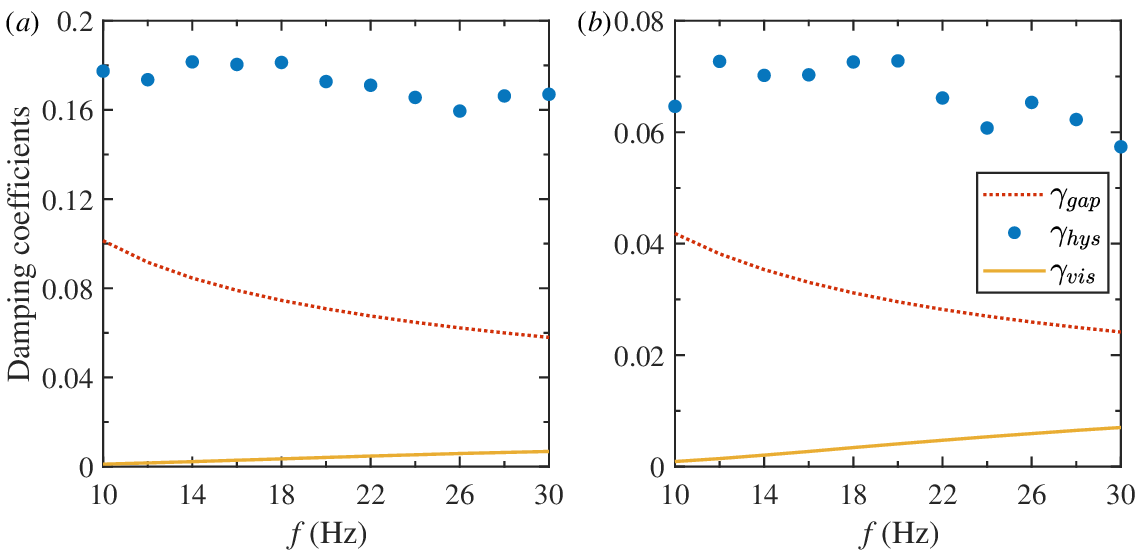}}
\caption{Damping coefficients with different origins versus driving frequency $f$ for 70$\%$ ethanol solution.
Dotted lines: the gap-averaged damping; solid dots: contact angle hysteresis dissipation; solid lines: bulk viscous dissipation.
The gap size is $b=2$ mm for (\textit{a}), and $b=4.8$ mm for (\textit{b}).
}
\label{fig:damping coefficients}
\end{figure}

In parametric excitation systems, Faraday instability emerges when the external forcing exceeds a critical threshold, disrupting the equilibrium between energy injection and dissipative losses.
In other words, the onset threshold is fundamentally determined by the dissipation, which, in the Hele-Shaw system, includes the gap-averaged damping resulting from the lateral walls, contact angle hysteresis losses at the pinned contact line, two-dimensional viscous dissipation in the fluid bulk, and end-wall boundary effects unique to finite-length rectangular containers.
The end-wall boundary effects are excluded from our discussion, since their impact on the instability problem is negligible, as validated by the experimental observations in figure~\ref{fig:different length and material}\textit{a},\textit{b}.
The remaining damping contributions are incorporated into the amplitude equations~\eqref{amplitude equation for faraday} and~\eqref{eq:non-dimen gap amplitude equation}, with their explicit expressions derived in~\eqref{eq:gamma1} and~\eqref{eq:gamma2}.
We isolate these damping coefficients according to their distinct physical origins, which read
\refstepcounter{equation}
\begin{equation}
\gamma_{gap} =- \frac{\delta_{1,i}}{2},\quad
\gamma_{hys} =\frac{2\mathrm{i}\sin\theta_s \rmDelta \sigma \kappa}{\pi \rho \Omega | \check{\eta}_{p1}'  |_{y'=b/2} \left | B' \right |},\quad
\gamma_{vis} =\delta_{St}^2k^2b^2.
\label{eq:damping coefficients}
\tag{\theequation{\textit{a}--\textit{c}}}
\end{equation}
Subscripts \textit{gap}, \textit{hys}, and \textit{vis} in~\eqref{eq:damping coefficients} denote, respectively, the gap-averaged damping, contact angle hysteresis dissipation, and bulk viscous dissipation.
All damping coefficients in \eqref{eq:damping coefficients} are strictly real-valued, because every parameter they contain is real other than the purely imaginary factor $\kappa$.
Figures~\ref{fig:damping coefficients}\textit{a},\textit{b} present the calculated damping coefficients $\gamma_{gap}$, $\gamma_{hys}$, and $\gamma_{vis}$ versus the driving frequency for pure ethanol in Hele-Shaw cells with $b=2$ mm and $b=4.8$ mm, respectively.
It should be clarified that the results of $\gamma_{hys}$ in figure~\ref{fig:damping coefficients} are not strictly smooth, because discrete values of $| \check{\eta}_{p1}' |_{y'=b/2}$, $|B'|$, and $\kappa$ that are determined numerically or experimentally are required for the calculation process.
The different frequency dependencies of these damping coefficients collectively result in the upward trend of $a_c$ shown in figure~\ref{fig:critical acceleration}.

Looking at each damping coefficient separately, according to~\eqref{eq:damping coefficients} and the definition of $\delta_{1,i}$, we find that $\tanh\left [ \left ( 1+\mathrm{i} \right )/2\delta _1  \right ]\approx 1$ in current configurations.
Consequently, $\gamma_{gap}$ is essentially governed by the modified dimensionless thickness of the Stokes boundary layer $\delta_1=2\sqrt{\nu/\Omega}/b$, yielding $\gamma_{gap}\approx \delta_1/2$.
Hence, the relationships between $f$, $b$ and $\gamma_{gap}$ are scaled as $\gamma_{gap}\propto f^{-1/2}$, and $\gamma_{gap}\propto b^{-1}$, indicating that $\gamma_{gap}$ decreases with increasing driving frequency and greater $b$.
Indeed, the varying trend of $\gamma_{gap}$ in figure~\ref{fig:damping coefficients} looks similar to that reported by \citet{li2019stability} using the Poiseuille flow assumption.
Moreover, rewriting $\gamma_{gap}$ in dimensional form gives $\gamma_{gap}' \approx\sqrt{\nu \Omega}/b$, which closely resembles the parietal friction damping $\gamma_w = \sqrt{\nu \Omega}/\sqrt{2}b$ proposed by \citet{cavelier2022subcritical}, with the factor $1/\sqrt{2}$ absorbed here because of the modified boundary layer thickness $\delta_1$.

Then, $\gamma_{hys}$ is examined.
According to~\eqref{eq:damping coefficients}, with fixed contact angle parameters, $\gamma_{hys}$ is determined by $\kappa$, $| \check{\eta}_{p1}' |_{y'=b/2}$, and $|B'|$.
According to the numerical results in Appendix~\ref{appen:Numerical solutions at order epsilon0}, we find that both magnitudes of $\kappa$ and $| \check{\eta}_{p1}' |_{y'=b/2}$ decrease linearly with increasing $f$.
The trend of $\gamma_{hys}$ is then determined by increasing $\Omega$ and decreasing $|B'|$.
The variation of these parameters with $f$ collectively captures the modifications of the meniscus dynamics under different forcing conditions.
If Hamraoui's model~\eqref{Hamraoui model} is used, according to \citet{bongarzone2023revised} and \citet{li2024stability}, $\gamma_{hys}$ is replaced by $\gamma_{ham}=2k\beta/\rho b \Omega$.
\citet{li2019stability} observed that $\gamma_{ham}$ increases monotonically with $f$, which ultimately leads to the overestimated trend of $a_c$ in figure~\ref{fig:critical acceleration}.
This discrepancy arises naturally due to the limitations of the gap-averaging methodology and Hamraoui's linear contact angle model.

Finally, the bulk viscous dissipation $\gamma_{vis}$ is discussed.
While viscous effects on the Faraday instability have been extensively investigated in three-dimensional systems (e.g., \citet{kumar1994parametric}; \citet{beyer1995faraday}; \citet{chen1999amplitude}), analogous studies in Hele-Shaw configurations are limited to the viscous theory developed by \citet{li2024stability}.
This initial theoretical framework established the complete viscous governing equations, but the explicit expression for viscous damping remains to be derived.
Fortunately, a viscous damping coefficient is obtained in~\eqref{eq:damping coefficients}, which is simplified to $\gamma_{vis}=2\nu k^2/\Omega$.
Given the linear relationship $k\propto f$ satisfied in the current range of $f$, we find $\gamma_{vis}\propto f$.
We also note that $\gamma_{vis}$ is independent of $b$, and its dimensional form $\gamma_{vis}'=2\nu k^2$ matches the classic expressions of \citet{lamb1945} and \citet{landau2013fluid}, as well as the three-dimensional results derived by \citet{kumar1994parametric} and \citet{chen1999amplitude}.
The key distinction lies in the replacement of the two-dimensional wavenumber $k$ with its three-dimensional counterpart, that is, the wavevector $\boldsymbol{k}=(k_x,k_y)$, where $k_x$ and $k_y$ represent the component in the $x'$ and $y'$ directions, respectively.
The effect of width is manifest for sufficiently large $b$ that Faraday waves exhibit transverse wave structures across the gap, namely $k_y\ne0$.
This phenomenon is not observed in Hele-Shaw experiments, and consequently $\gamma_{vis}$ is independent of $b$.
Furthermore, figure~\ref{fig:damping coefficients} indicates that, compared to the other two damping terms, the viscous damping is negligible under present experimental conditions, justifying its ignorance as adopted in several previous studies \citep{rajchenbach2011new,li2019stability,bongarzone2023revised}.
Nevertheless, its influence will become appreciable if the fluid viscosity is prominent.
In such configurations, viscous dissipation could no longer be neglected, a point that merits further experimental evidence.

Before concluding, we give a brief discussion on how key physical parameters (e.g., kinematic viscosity $\nu$, surface tension coefficient $\sigma$, static contact angle $\theta_s$, and contact angle hysteresis range $\rmDelta$) influence damping in the Hele-Shaw system.
According to~\eqref{eq:damping coefficients}, the first parameter $\nu$ enters the damping through $\gamma_{gap}$ and $\gamma_{vis}$, with scale relationships as $\gamma_{gap}\propto \nu^{1/2}$ and $\gamma_{vis}\propto \nu$.
We note that the gap-averaged damping also arises from fluid viscosity, as it is induced by the no-slip condition at lateral walls, where the viscous Stokes boundary layer exists.
However, one should distinguish it from the so-called viscous dissipation, which originates from the fluid bulk, as clearly delineated by \citet{li2024stability}.
The other parameters $\sigma$, $\theta_s$, and $\rmDelta$ act directly on $\gamma_{hys}$, as they are all characterized on the free surface.
In particular, $\sigma$ may additionally affect $\gamma_{vis}$ indirectly through its impact on $k$ in the dispersion relation.
Excluding this potential influence, $\gamma_{hys}$ exhibits a linear dependence on $\sigma$, scaling as $\gamma_{hys}\propto \sigma$.
In terms of wetting parameters, we have examined that despite the minimal influence of $\theta_s$ on $\kappa$ through static state modifications, the contact angle hysteresis dissipation is enhanced for lager $\theta_s$, with an approximate trend $\gamma_{hys}\propto \sin \theta_s $.
For $\rmDelta$, there is an evident relationship in~\eqref{eq:damping coefficients} that $\gamma_{hys}\propto\rmDelta$, namely the contact angle damping is proportional to the hysteresis range.

\section{Concluding remarks}
\label{sec:Concluding remarks}

\citet{li2019stability} confirmed that the effect of the dynamic contact angle dominates over the confinement of the lateral walls in governing Faraday instability problem in Hele-Shaw cells, revealing fundamentally different dynamics from the configuration of wide-mouth containers \citep{milner1991square}.
However, existing theoretical frameworks typically rely on the gap-averaged governing equations combined with Hamraoui's linear dynamic contact angle model \citep{bongarzone2023revised,li2024stability}.
Experimental observations of the meniscus suggest that the contact line remains pinned while the meniscus oscillates periodically before Faraday onset emerges, demonstrating that Hamraoui's model results in misrepresented meniscus dynamics and exhibits significant contradictions in oscillatory fluid systems, ultimately yielding inaccurate damping evaluation.

In this paper, we present a gap-resolved approach that directly deals with the transverse gap flow in Hele-Shaw cells, thereby avoiding oversimplifying contact angle dynamics.
A refined contact angle hysteresis model is implemented in boundary conditions at lateral walls, accurately capturing capillary hysteresis in meniscus dynamics.
By means of an asymptotic expansion technique, an amplitude equation for the meniscus that explicitly accounts for the contact angle hysteresis damping is derived from governing equations for the gap flow.
This damping depends on contact angle parameters (e.g. static contact angle and hysteresis range), numerical solutions and experimental measurements characterizing meniscus dynamics, and external forcing frequency.
A novel amplitude equation for linear Faraday instability is developed by assuming an oscillatory Stokes flow profile and incorporating contact angle hysteresis damping.
Linear stability analysis is conducted by using Lyapunov's first method.
Comparison with experiments performed in a wide driving frequency range validates the present theory for addressing the Faraday instability problem in Hele-Shaw cells.
This improvement stems from two key factors: the gap-resolved treatment of transverse gap flow and precise measurements of contact angle parameters through meniscus experiments in a container with material properties identical to those used in Faraday instability experiments.

Through the asymptotic analysis, contact angle hysteresis damping effectively modifies the frequency shift, thereby partially compensating for the detuning that has persisted since the introduction of Stokes boundary layer theory.
% Such detuning compensation is, however, not pronounced in the Floquet analysis.
% Therefore, the asymptotic expansion approach offers specific advantages for the present linear instability problem: by relegating small contributions to higher orders, it effectively captures the modifications to the dispersion relation that arise from hysteresis damping and viscous dissipation, while yielding onset thresholds that differ negligibly from those obtained via Floquet analysis.}
Nevertheless, since the formation of a liquid film during experimental measurements of the critical wavenumber indicates a significant change in the wettability condition, and nonlinear features of Faraday waves become more prominent beyond the onset, the linear stability analysis is not strictly appropriate to explain the dispersion relation of Faraday waves in Hele-Shaw cells.
Moreover, our analysis reveals that contact angle hysteresis damping exhibits a trend governed by competing effects of external forcing and meniscus dynamics, and demonstrates scaling relationships with contact angle parameters, specifically $\gamma_{hys}\propto \sin \theta_s$ and $\gamma_{hys}\propto \rmDelta$.
The gap-averaged damping is quantitatively determined by the dimensionless thickness of the Stokes boundary layer, with a scaling as $\gamma_{gap}\propto \sqrt{\nu/\Omega}/b$.
Furthermore, we derive for the first time the viscous dissipation of Faraday waves in Hele-Shaw cells, formulated as $\gamma_{vis}=2\nu k^2/\Omega$, which maintains a form similar to the one in wide-mouth containers.

Although the established final relation~\eqref{eq:ak} between the critical value of onset and parameters characterizing each physical process is theoretically self-contained and has been verified to be a well-matched model to reveal the generation mechanism and evolution of Faraday instability in Hele-Shaw cells,
the validation strongly depends on the laboratory measurements of these parameters, which has been well demonstrated in the present study.
Among them, only $\left | B' \right |$ is obtained as the a posteriori value from the experiment, although $\left | B' \right |$ is measured before the onset from a stable signal recognized out of the raw data right before the rapid rise of the free surface.
Therefore, in practice, the dependence of $\left | B' \right |$ on other a priori ones becomes a must before utilizing Equation~\eqref{eq:ak} to predict the critical acceleration.
For a possible theoretical approach, \citet{bongarzone2022subharmonic} proposed a solution procedure: combining the $\textit{O} (\epsilon)$ driving acceleration with the static meniscus produces a non-resonant harmonic forcing term proportional to the acceleration; solving for the linear response to this forcing would yield an inviscid meniscus wave solution.
This route provides a promising direction for research.

% The combination of theoretical and experimental approaches elucidates the generation mechanism and evolution of Faraday instability in Hele-Shaw cells, which are strongly influenced by contact angle dynamics.
% The critical condition for the onset is primarily governed by oscillatory Stokes flow and meniscus dynamics.
% However, to complete the calculation of the onset thresholds, the proposed theory relies heavily on experimentally determined values of $\left | B' \right |$ that are measured from the oscillation amplitude of the meniscus bottom in the stable stage immediately preceding the rapid rise of the free surface.
% Although $\left | B' \right |$ is an experimental parameter measured before the onset, its mathematical relationships with external forcing and wettability must be clarified to accomplish a fully theoretical prediction.
% With the theoretically estimated $\left | B' \right |$, a prior linear stability analysis becomes more achievable.

\begin{bmhead}[Supplementary material]
The supplementary material presents the image analysis procedure used in \S~\ref{sec:Meniscus observation} to extract the dynamic contact angle, a Floquet analysis of the full gap-averaged model mentioned in \S~\ref{sec:Dispersion relation}, and a table of symbols clarifying their physical interpretation.
\end{bmhead}

\begin{bmhead}[Acknowledgements]
We thank the anonymous reviewers for helping strengthen the manuscript and Runsheng Chen for help with the laboratory experiments.
\end{bmhead}

\begin{bmhead}[Funding]
This work was supported by the National Natural Science Foundation of China (Approval No. 12002206) and the Fundamental Research Funds for the Central Universities.
\end{bmhead}

\begin{bmhead}[Declaration of interests]
The authors report no conflict of interest.
\end{bmhead}

%\appendix
\begin{appen}

\section{Gap-averaged asymptotic analysis}
\label{app:Gap-averaged asymptotic analysis}

\subsection{Asymptotic solutions of Faraday instability}
\label{app:Asymptotic solutions of Faraday instability}

In \S~\ref{sec:Faraday asymptotic expansion}, the expanded governing equations~\eqref{rescaled laplace}--\eqref{rescaled bottom condition} are solved analytically.
For each order problem,~\eqref{rescaled laplace} and~\eqref{rescaled bottom condition} will not change, and thus only boundary conditions on the free surface are listed.
At order $\epsilon^0$,~\eqref{rescaled normal stress condition} and~\eqref{rescaled kinematic condition} reduce to
\begin{subequations}
\begin{equation}
-\bar{p}_1+ \bar{\zeta}_1 -l_c^2 k^2\frac{\partial^2 \bar{\zeta}_1}{\partial x^2} =0 ,\quad \text{at }z=0,
\label{O1 rescaled normal stress condition}
\end{equation}
\begin{equation}
4 \omega_0^2 \left ( 1+l_c^2k^2 \right ) \frac{\partial \bar{\zeta}_1}{\partial t} -2\mathrm{i} \frac{\partial \bar{p}_1}{\partial z}=0,\quad \text{at }z=0.
\label{O1 rescaled kinematic condition}
\end{equation}
\end{subequations}
The solutions read
\begin{subequations}
\begin{equation}
\bar{p}_1=\left ( 1+l_c^2k^2 \right )\cos xe^ze^{\mathrm{i}t/2 }A\left ( T \right )+\text{c.c.},
\label{O1 solution for p}
\end{equation}
\begin{equation}
\bar{\zeta}_1=\cos xe^{\mathrm{i}t/2 }A\left ( T \right )+\text{c.c.},
\label{O1 solution for zeta}
\end{equation}
\end{subequations}
where $\omega_0=1$ represents a traditional dispersion relation for gravity-capillary waves, $A(T)$ is the complex amplitude of the Faraday wave on a slow time scale, and c.c. denotes the complex conjugate.

At order $\epsilon$,~\eqref{rescaled normal stress condition} and~\eqref{rescaled kinematic condition} are written as
\begin{subequations}
\begin{equation}
-\bar{p}_2+ \bar{\zeta}_2-l_c^2 k^2 \frac{\partial^2 \bar{\zeta}_2}{\partial x^2}=\frac{\hat{a}}{g}\cos t  \bar{\zeta}_1-2\mathrm{i}\hat{\delta} \frac{\partial^2 \bar{p}_1}{\partial z^2} ,\quad \text{at }z=0,
\label{O2 rescaled normal stress condition}
\end{equation}
\begin{equation}
4\left ( 1+l_c^2 k^2 \right )  \frac{\partial \bar{\zeta}_2}{\partial t}-2\mathrm{i} \frac{\partial \bar{p}_2}{\partial z}=-4\left ( 1+l_c^2 k^2 \right ) \frac{\partial \bar{\zeta}_1}{\partial T} -4\omega_1^2\left ( 1+l_c^2 k^2 \right ) \frac{\partial \bar{\zeta}_1}{\partial t} -2\mathrm{i}\hat{\delta} _1 \frac{\partial \bar{p}_1}{\partial z} ,\quad \text{at }z=0.
\label{O2 rescaled kinematic condition}
\end{equation}
\end{subequations}
By solving $\bar{\zeta_2}$ using~\eqref{O2 rescaled kinematic condition}, and combining it with~\eqref{O2 rescaled normal stress condition}, an equation of $\bar{p}_2$ is obtained.
Substituting~\eqref{O1 solution for p} and~\eqref{O1 solution for zeta} into this equation, according to the Fredholm alternative, a solvability condition appears on the right-hand side, resulting in the amplitude equation for the linear Faraday instability~\eqref{amplitude equation for faraday}.

\subsection{Linear stability analysis based on Hamraoui's model}
\label{Linear stability analysis based on Hamraoui's model}

In \S~\ref{sec:Incorporating meniscus effects into gap-averaged equations}, we have briefly discussed how Hamraoui's model~\eqref{Hamraoui model} can be used to incorporate meniscus effects into gap-averaged equations, along with its limitations in addressing the Faraday instability problem in Hele-Shaw cells.
In this appendix, we continue the derivation starting from~\eqref{eq:gap dimensionless normal stress condition} and establish the full gap-averaged model based on Hamraoui's model while retaining viscous effects.
Subsequently, Lyapunov's first method is employed to obtain the critical condition for Faraday onset.
These results help clarify the influences of contact angle dynamics on the Faraday instability.

Substituting the contact angle condition based on Hamraoui's model~\eqref{Hamraoui boundary} into~\eqref{eq:gap dimensionless normal stress condition} yields the gap-averaged normal stress boundary condition, with contact angle dynamics and viscous effects retained:
\begin{equation}
-\bar{p}+\left ( 1- \frac{a}{g}\cos t \right ) \bar{\zeta}+\delta _{St}^2k^2b^2\lambda _1\frac{\partial^2 \bar{p}}{\partial z^2}=l_c^2 \left ( k^2\frac{\partial^2 \bar{\zeta}}{\partial x^2}-\frac{2k\beta g}{b\sigma \Omega} \lambda_1 \frac{\partial \bar{p}}{\partial z} \right ),\quad \text{at }z=0,
\label{aqq:eq:gap dimensionless normal stress condition}
\end{equation}
which, together with~\eqref{laplace equation},~\eqref{non penetrable condition}, and~\eqref{averaged kinematic boundary}, constitutes the full gap-averaged governing equations.

Following the same asymptotic approach as used in \S~\ref{sec:Faraday asymptotic expansion}, another amplitude equation is derived:
\begin{equation}
\frac{\mathrm{d} A}{\mathrm{d} T} +\frac{1}{2} \left [ \mathrm{i} \left ( \hat{\delta}_{1,r}+\mathrm{i} \hat{\delta}_{1,i} \right )+\mathrm{i}\omega_1^2+2\hat{\delta}+2\hat{\beta} \right ] A+\frac{\mathrm{i}\hat{a}}{4g\left ( 1+l_c^2k^2 \right ) }A^*=0,
\label{app:eq:amplitude equation}
\end{equation}
where $\hat{\beta}$, defined by $\epsilon \hat{\beta}=2k\beta g l_c^2/ b\sigma \Omega$, is a rescaled parameter associated with contact angle effects, consistent with the scaling introduced in~\eqref{parameter expanding of Faraday}.
Compared with~\eqref{eq:combined amplitude equation}, the damping term $\chi$ is here replaced by $\hat{\beta}$.
Applying Lyapunov's first method outlined in \S~\ref{sec:Stability analysis}, an analytical expression for the critical acceleration is obtained:
\begin{equation}
a_k^{Ham}= 2g \left (1+l_c^2k^2  \right )\sqrt{\left ( 2\delta _{St}^2k^2b^2 +\frac{4k\beta }{\rho b \Omega} -\delta_{1,i} \right )^2 +\left ( \delta_{1,r} +\omega^2-1\right ) ^2}.
\label{eq:ak_ham}
\end{equation}

\section{Gap-resolved transverse flow}
\label{appen:Gap-resolved transverse flow}

\subsection{Static meniscus}
\label{appen:Static meniscus}

Governing equations for the static meniscus~\eqref{eq:phi_s=0}--\eqref{eq:theta_s} are solved by a numerical approach \citep{viola2018capillary}.
Equation~\eqref{eq:eta_s} is discretized by means of the Chebyshev collocation method on the Gauss--Lobatto--Chebyshev collocation grid $s\in [-1,1]$, which is mapped into the physical space $y'\in [0,b/2]$ (one can extend to the entire domain region according to symmetry) through a linear mapping $y'=(s+1)b/4$.
The differential operators in~\eqref{eq:eta_s} and~\eqref{eq:theta_s} are then represented by algebraic matrices.
The iterative Newton method includes five steps:
\vspace{6pt}
\begin{enumerate}[label=(\roman*)]
\item Define an initial guess solution $\eta_s'^{(0)}$ that satisfies the boundary condition~\eqref{eq:theta_s}.
\item Calculate the residual function $F ( \eta_s'^{(i)} )$ and the corresponding Jacobian $J ( \eta_s'^{(i)} )=dF ( \eta_s'^{(i)} )/d \eta_s'^{(i)}$.
\item Solve for $\delta \eta_s'$ the linear system $J ( \eta_s'^{(i)} ) \delta \eta_s'=- F ( \eta_s'^{(i)} )$.
\item Update the solution $\eta_s'^{(i+1)}=\eta_s'^{(i)}+\delta \eta_s'$.
\item Compute the $L_2$-norm of $\delta \eta_s'$.
If $\left \| \delta \eta_s' \right \|< 10^{-8}$, output the solution for $\eta_s'$, otherwise, go back to step (ii).
\end{enumerate}
\vspace{6pt}

\subsection{Numerical solutions of $\check{\phi}_{p1}'$ and $\check{\eta}_{p1}'$ at order $\epsilon^0$}
\label{appen:Numerical solutions at order epsilon0}

In \S~\ref{sec:Order epsilon^0}, the solutions at order $\epsilon^0$ are obtained numerically.
The dynamic condition~\eqref{eq:gap dynamic condition} at this order reads
\begin{equation}
\frac{\partial \phi_{p1}'}{\partial t'} + g \eta_{p1}' - \frac{\sigma}{\rho} \left \{ \frac{\partial _{y'y'}\eta_{p1}' }{\left [ 1+\left (\partial _{y'}\eta_s'  \right ) ^2 \right ]^{3/2} } + \frac{\partial _{x'x'}\eta_{p1}' }{\left [ 1+\left (\partial _{y'}\eta_s'  \right ) ^2 \right ]^{1/2} } \right \} =0,\quad \text{at }z'=0.
\label{eq:gap O1 dynamic condition}
\end{equation}
Remaining equations for $\phi_{p1}'$ and $\eta_{p1}'$ that have the same form as~\eqref{eq:gap laplace equation}, and~\eqref{eq:gap kinematic condition}--\eqref{eq:gap lateral condition} are not listed again.

Substituting the ansatzes~\eqref{eq:ansatz of phi1} and~\eqref{eq:ansatz of eta1} into the governing equations at this order, one can obtain a system of equations about the spatial variables $\check{\phi}_{p1}'$ and $\check{\eta}_{p1}'$, whose form in the generalized eigenvalue equation reads
\begin{equation}
\left ( \mathrm{i} \Omega \mathsfbi{B} -\mathsfbi{A} \right )  \check{\boldsymbol{q}}_1'=\boldsymbol{0},
\label{eq:gap O1 state euqations}
\end{equation}
where $\check{\boldsymbol{q}}_1'=(\check{\phi}_{p1}',\check{\eta}_{p1}')^\mathrm{T}$ is the eigenvector, and the linear operators are defined by
\begin{equation}
\mathsfbi{B} = \begin{pmatrix} 0 & 0 \\ \boldsymbol{I} & 0 \end{pmatrix},\quad  \mathsfbi{A} = \begin{pmatrix} -k_m ^2+\frac{\partial^2 }{\partial y'^2 } +\frac{\partial^2 }{\partial z'^2 } & 0 \\ 0 & -g\boldsymbol{I}+ \frac{\sigma}{\rho} \left \{ \frac{\partial _{y'y'} }{\left [ 1+\left (\partial _{y'}\eta_s'  \right ) ^2 \right ]^{3/2} }- \frac{k_m ^2 }{\left [ 1+\left (\partial _{y'}\eta_s'  \right ) ^2 \right ]^{1/2} } \right \} \end{pmatrix}.
\label{eq:gap O1 matrices expression}
\end{equation}
Equation~\eqref{eq:gap O1 state euqations} is restricted by the following boundary conditions:
\refstepcounter{equation}
\begin{equation}
\mathrm{i} \Omega \check{\eta}_{p1}'-\frac{\partial \check{\phi}_{p1}'}{\partial z'} + \frac{\partial \eta_s'}{\partial y'} \frac{\partial \check{\phi}_{p1}'}{\partial y'} = 0,\quad \text{at }z'=0,
\tag{\theequation{\textit{a}}}
\label{eq:gap O1 free surface condition}
\end{equation}
\begin{equation}
\left. \frac{\partial \check{\phi}_{p1}'}{\partial z'} \right |_{z'=-H} =0,\qquad
\left. \frac{\partial \check{\phi}_{p1}'}{\partial y'} \right |_{y'=\pm b/2} =0,\qquad
\left. \frac{\partial \check{\eta}_{p1}'}{\partial y'} \right |_{y'=\pm b/2} =0.
\tag{\theequation{\textit{b}--\textit{d}}}
\label{eq:gap O1 lateral conditions}
\end{equation}

The coupled system of the eigenvalue equation and boundary conditions is solved numerically by means of a spectral method \citep{viola2016mode,viola2018capillary}.
For the values of $k_m $, we adopt the results calculated from the classical dispersion relation for gravity-capillary waves, namely $\Omega^2=g k_m+\sigma k_m^3/\rho$.
Linear operators $\mathsfbi{B}$ and $\mathsfbi{A}$ are discretized using the Chebyshev collocation method.
Differential operators in~\eqref{eq:gap O1 free surface condition} and~\eqref{eq:gap O1 lateral conditions} are also discretized via spectral differentiation matrices, and then replace the corresponding rows of the algebraic matrices $\mathsfbi{A}$ and $\mathsfbi{B}$.
Eventually,~\eqref{eq:gap O1 state euqations} is solved in MATLAB for the eigenvector $\check{\boldsymbol{q}}_1'$.
The eigenvalue in~\eqref{eq:gap O1 state euqations} corresponds to the response frequency and ultimately leads to a theoretical dispersion relation at this order, although a detailed discussion is not provided here.
A grid resolution of $N_y=N_z=60$ nodes is employed, which ensures the numerical convergence of the results.
For verification of grid convergence, we present in figure~\ref{fig:convergence analysis} the numerical results of $-\Imag(\kappa)$ (with $\kappa$ defined in~\eqref{eq:gap amplitude equation}) obtained using a finer grid $N_y=N_z=80$.
Without loss of generality, the solution of $\check{\boldsymbol{q}}_1'$ is normalized by setting $\left | \check{\eta}_{p1}' \right |_{y'=0} =1$.

\begin{figure}
\centerline{\includegraphics[width=10cm]{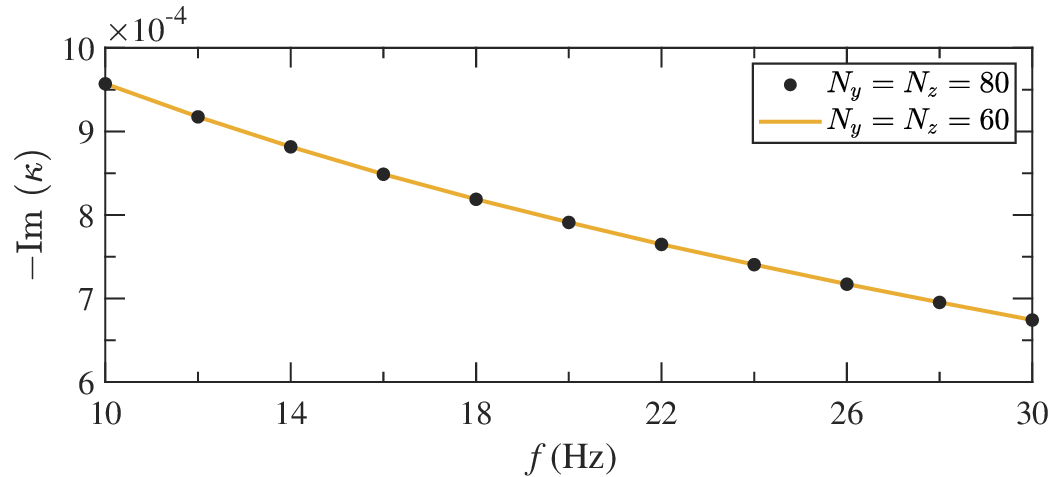}}
\caption{Convergence analysis for the $\epsilon^0$-order numerical implementation.
Results show the imaginary part of $\kappa$ calculated by~\eqref{eq:gap amplitude equation} against driving frequency $f$.
The solid line is computed using a $N_y=N_z=60$ computational grid, while the dots are obtained with a finer $N_y=N_z=80$ computational grid.
Physical parameters of pure ethanol are used as inputs, while $b=2$ mm.}
\label{fig:convergence analysis}
\end{figure}

\subsection{Solvability condition at order $\epsilon$}
\label{appen:Solvability condition for the gap flow}

In \S~\ref{sec:Order epsilon}, an amplitude equation is derived from the solvability condition at order $\epsilon$.
At this order, the form of~\eqref{eq:gap laplace equation},~\eqref{eq:gap bottom condition}, and~\eqref{eq:gap lateral condition} will not change, while~\eqref{eq:gap dynamic condition},~\eqref{eq:gap kinematic condition}, and~\eqref{eq:gap geometrical relation} read
\begin{subequations}
\begin{equation}
\frac{\partial \phi_{p2}'}{\partial t'} + g\eta_{p2}' - \frac{\sigma}{\rho} \left \{ \frac{\partial _{y'y'}\eta_{p2}' }{\left [ 1+\left (\partial _{y'}\eta_s'  \right ) ^2 \right ]^{3/2} }+ \frac{\partial _{x'x'}\eta_{p2}' }{\left [ 1+\left (\partial _{y'}\eta_s'  \right ) ^2 \right ]^{1/2} } \right \}=- \frac{\partial \phi_{p1}'}{\partial T'},\quad \text{at }z'=0,
\label{eq:gap O2 dynamic condition}
\end{equation}
\begin{equation}
\frac{\partial \eta_{p2}'}{\partial t'} + \frac{\partial \eta_s'}{\partial y'} \frac{\partial \phi_{p2}'}{\partial y'} - \frac{\partial \phi_{p2}'}{\partial z'} = - \frac{\partial \eta_{p1}'}{\partial T'},\quad \text{at }z'=0,
\label{eq:gap O2 kinematic condition}
\end{equation}
\begin{equation}
\frac{\partial \eta_{p2}'}{\partial y'} =\mp \frac{ \theta_{p2}}{\sin^2 \theta_s } ,\quad \text{at } y'=\pm \frac{b}{2}.
\label{eq:gap O2 contact line condition}
\end{equation}
\end{subequations}
Since we are only concerned with the expression for the capillary damping, which will be incorporated into~\eqref{amplitude equation for faraday} for the stability analysis, the external forcing term $\hat{a}\cos \Omega t' \eta_{p1}'$ is not explicitly written in~\eqref{eq:gap O2 dynamic condition}.
Substituting the ansatzes~\eqref{eq:gap O2 ansatz of phi2} and~\eqref{eq:gap O2 ansatz of eta2} into the governing equations, an equation similar to~\eqref{eq:gap O1 state euqations} is obtained:
\begin{equation}
\left ( \mathrm{i} \Omega \mathsfbi{B} -\mathsfbi{A} \right )  \tilde{ \boldsymbol{q} }_2=\check{\boldsymbol{D}}_1',
\label{eq:gap O2 state equations}
\end{equation}
with
\refstepcounter{equation}
\begin{equation}
\tilde{ \boldsymbol{q} }_2=\begin{pmatrix} \tilde{\phi}_{p2}' \\ \tilde{\eta}_{p2}' \end{pmatrix},\quad 
\check{\boldsymbol{D}}_1'=\begin{pmatrix} 0 \\ -\frac{\mathrm{d} B'}{\mathrm{d} T'} \check{\phi}'_{p1} \end{pmatrix}.
\tag{\theequation{\textit{a},\textit{b}}}
\label{eq:gap o2 matric expressions}
\end{equation}
Equation~\eqref{eq:gap O2 state equations} is subject to the contact line condition~\eqref{eq:gap O2 contact line condition} and
\refstepcounter{equation}
\begin{equation}
\mathrm{i} \Omega \tilde{\eta}_{p2}' -\frac{\partial \tilde{\phi}_{p2}'}{\partial z'} + \frac{\partial \eta_s'}{\partial y'} \frac{\partial \tilde{\phi}_{p2}'}{\partial y'} =-\frac{\mathrm{d} B'}{\mathrm{d} T'}\check{\eta}_{p1}',\quad \text{at } z'=0,
\tag{\theequation{\textit{a}}}
\label{eq:gap O2 no time kinematic condition}
\end{equation}
\begin{equation}
\left. \frac{\partial \tilde{\phi}_{p2}'}{\partial z'} \right |_{z'=-H}=0,\qquad \left. \frac{\partial \tilde{\phi}_{p2}'}{\partial y'} \right |_{y'=\pm b/2} =0.
\tag{\theequation{\textit{b},\textit{c}}}
\label{eq:gap O2 no time lateral condition}
\end{equation}

To obtain the solvability condition, the adjoint global mode $ \boldsymbol{q}_1^\dagger =(\phi_{p1}^\dagger, \eta_{p1}^\dagger)^\mathrm{T} $ is introduced, which is the solution of the adjoint equation:
\begin{equation}
\left ( \mathrm{i} \Omega \mathsfbi{B}^\dagger -\mathsfbi{A}^\dagger \right )  \boldsymbol{q}_1^\dagger=\boldsymbol{0},
\label{eq:gap O2 adjoint equations}
\end{equation}
where the linear operators $\mathsfbi{A}^\dagger$ and $\mathsfbi{B}^\dagger$ are derived by integrating by parts the system of~\eqref{eq:gap O1 state euqations}.
Actually, the direct problem at order $\epsilon^0$ is self-adjoint with respect to the Hermitian scalar product, which means
\begin{equation}
\mathrm{i} \Omega \mathsfbi{B}^\dagger -\mathsfbi{A}^\dagger = -\mathrm{i} \Omega \mathsfbi{B} -\mathsfbi{A}.
\label{eq:gap O2 adjoint operator}
\end{equation}
Therefore, the complex conjugate of the adjoint mode satisfies $\phi_{p1}^{\dagger*}=\check{\phi}_{p1}'$ and $\eta_{p1}^{\dagger*}=\check{\eta} _{p1}'$.
The interested readers can refer to \citet{viola2018capillary} for the complete derivation of the adjoint equation and the definition of the adjoint mode.

According to the Fredholm alternative,~\eqref{eq:gap O2 state equations} has a nontrivial solution if and only if $\check{\boldsymbol{D}}_1'$ is orthogonal to the adjoint mode $\boldsymbol{q}_1^\dagger$, which leads to
\begin{equation}
\left \langle \boldsymbol{q}_1^\dagger,\left (\mathrm{i}\Omega \mathsfbi{B} -\mathsfbi{A} \right ) \tilde{ \boldsymbol{q} }_2 \right \rangle = \left \langle \boldsymbol{q}_1^\dagger, \check{\boldsymbol{D}}_1'  \right \rangle.
\label{eq:gap O2 compatibility condition}
\end{equation}
The brackets $\left \langle  \right \rangle$ defines the Hermitian scalar product:
\begin{equation}
\left \langle \boldsymbol{q}_\alpha ,\boldsymbol{q}_{\vartheta}  \right \rangle =\int _V \phi _{\alpha}^*\phi_{\vartheta}\mathrm{d} V +\int_{-b/2}^{b/2}\eta_{\alpha}^*\eta_{\vartheta }\mathrm{d}y'.
\label{eq:gap Hermitian production}
\end{equation}
By substituting the expression of $\check{\boldsymbol{D}}_1'$ defined in~\eqref{eq:gap o2 matric expressions}, the right-hand side of~\eqref{eq:gap O2 compatibility condition} reads
\begin{equation}
\left \langle \boldsymbol{q}_1^\dagger, \check{\boldsymbol{D}}_1'  \right \rangle =-\frac{\mathrm{d} B'} {\mathrm{d} T'} \int_{-b/2}^{b/2} \eta_{p1} ^{\dagger*} \left. \check{\phi}_{p1}' \right |_{z'=0}\mathrm{d} y'.
\label{eq:gap O2 right side compatibility}
\end{equation}
Furthermore, by combining with the boundary conditions~\eqref{eq:gap O2 contact line condition},~\eqref{eq:gap O2 no time kinematic condition}, and~\eqref{eq:gap O2 no time lateral condition}, the left-hand side of~\eqref{eq:gap O2 compatibility condition} is expanded as
\begin{eqnarray}
\left \langle \boldsymbol{q}_1^\dagger,\left ( \mathrm{i}\Omega \mathsfbi{B} -\mathsfbi{A} \right ) \tilde{ \boldsymbol{q} }_2 \right \rangle = \left \langle \left ( \mathrm{i}\Omega \mathsfbi{B}^\dagger -\mathsfbi{A}^\dagger  \right )\boldsymbol{q}_1^\dagger,\tilde{ \boldsymbol{q} }_2 \right \rangle + \frac{\mathrm{d} B'}{\mathrm{d} T'} \int _{-b/2}^{b/2} \check{\eta}'_{p1} \left. \phi_{p1} ^{\dagger*}\right |_{z'=0}\mathrm{d}y' \nonumber \\ - \frac{2\sigma\sin^3\theta_s}{\rho}\eta_{p1} ^{\dagger*}  \left. \frac{\partial \tilde{\eta} _{p2}'}{\partial y'} \right |_{y'=b/2}.
\label{eq:gap O2 left side compatibility}
\end{eqnarray}
On the right-hand side of~\eqref{eq:gap O2 left side compatibility}, the first term is zero because of the definition of adjoint equation~\eqref{eq:gap O2 adjoint equations}, the second term comes from the slow time derivative in the kinematic condition~\eqref{eq:gap O2 no time kinematic condition}, and the last term is associated with the contact line condition~\eqref{eq:gap O2 contact line condition}.
The solvability condition~\eqref{eq:gap O2 compatibility condition} then reduces to
\begin{equation}
\frac{\mathrm{d} B'}{\mathrm{d} T'}\int _{-b/2}^{b/2}\left. \left (  \eta_{p1} ^{\dagger*} \check{\phi}'_{p1}+\phi_{p1} ^{\dagger*}\check{\eta}'_{p1} \right )\right |_{z'=0}\mathrm{d} y'-\frac{2\sigma\sin^3\theta_s}{\rho}\eta_{p1} ^{\dagger*} \left. \frac{\partial \tilde{\eta} _{p2}'}{\partial y'} \right |_{y'=b/2}=0.
\label{eq:gap O2 final compatibility condition}
\end{equation}

To derive the amplitude equation from~\eqref{eq:gap O2 final compatibility condition}, the contact angle at this order~\eqref{eq:gap O2 theta_2} should be dealt with appropriately.
Now that the tangent function is nonlinear, a direct treatment is introducing the Fourier series expansion and separating out the subharmonic modes \citep{nayfeh1993introduction,viola2018capillary}:
\begin{equation}
\tanh \left ( \frac{\hat{\alpha}}{\epsilon} \left. \frac{\partial \eta_{p1}'}{\partial t'} \right |_{y'= b/2} \right )=\tanh \left [ \frac{\hat{\alpha}}{\epsilon} \mathrm{i}\Omega B' \left. \check{\eta}'_{p1} \right |_{y'=b/2}e^{\mathrm{i}( \Omega t'+k_m x')}+\text{c.c.}  \right ] =\sum_{n=-\infty }^{\infty } d_ne^{\mathrm{i} n \Omega t'}.
\label{eq:gap O2 fourier expansion}
\end{equation}
If we decompose the complex amplitude $B'$ and $\check{\eta}_{p1}'$ in modulus and phase, namely $B'\left(T' \right)=\left | B' \right |e^{\mathrm{i}\vartheta \left ( T' \right ) } $ and $\check{\eta}_{p1}'=\left | \check{\eta}_{p1}' \right |e ^{\mathrm{i}\vartheta _{\eta}}$, the coefficients $d_n$ are then given by
\begin{equation}
d_n=\frac{\Omega}{2\pi} \int_{-\pi/\Omega}^{\pi/\Omega }\tanh  \left ( \frac{\hat\alpha}{\epsilon}\Omega \left | B' \right | \left | \check{\eta}_{p1}' \right |_{y'=b/2} \cos \psi   \right ) e^{- \mathrm{i}n\Omega t'} \mathrm{d}t',
\label{eq:gap O2 dn intergrate}
\end{equation}
where the variable $\psi$ is defined as $\psi =\Omega t'+k_m x'+\pi/2+\vartheta \left ( T' \right )+\left. \vartheta _{\eta} \right |_{y'=b/2}$.
Since the rescaled steepness coefficient $\hat{\alpha}/\epsilon$ is large enough, the hyperbolic tangent can be expanded as the sum of the sign function plus a small correction $f(\epsilon,\psi )$, which satisfies $\int_{-\pi}^{\pi} f(\epsilon,\psi ) e^{-\mathrm{i}n\Omega t'} \mathrm{d} \psi = \textit{O}(\epsilon)$.
Hence, equation~\eqref{eq:gap O2 dn intergrate} is expanded as
\begin{eqnarray}
d_n&=&\frac{\Omega}{2\pi} \int_{-\pi/\Omega}^{\pi/\Omega} \mathrm{sgn} \left ( \frac{\hat\alpha}{\epsilon}\Omega \left | B' \right | \left | \check{\eta}_{p1}' \right |_{y'=b/2} \cos \psi  \right )e^{-\mathrm{i}n\Omega t'} \mathrm{d}t' + \textit{O} \left ( \epsilon  \right ) \nonumber \\
&=& e^{\mathrm{i}n\left ( k_m x'+\pi/2 + \vartheta \left ( T' \right ) + \left. \vartheta _{\eta} \right |_{y'=b/2} \right ) }\frac{1}{2\pi}  \int_{-\pi}^{\pi} \mathrm{sgn} \left ( \cos \psi \right ) e^{-\mathrm{i}n \psi } \mathrm{d}\psi + \textit{O} \left ( \epsilon  \right ) \nonumber \\
&=& \frac{2}{\pi}c_n e^{\mathrm{i}n\left ( k_m x'+\pi/2 + \vartheta \left ( T' \right ) + \left. \vartheta _{\eta} \right |_{y'=b/2} \right ) }+\textit{O} \left ( \epsilon  \right ).
\label{eq:gap O2 dn expanded}
\end{eqnarray}
When $n=1$, the Fourier coefficient $c_1$ in~\eqref{eq:gap O2 dn expanded} reads
\begin{equation}
\frac{2}{\pi}c_1=\frac{1}{2\pi}  \int_{-\pi}^{\pi} \mathrm{sgn} \left ( \cos \psi \right ) e^{-\mathrm{i} \psi } \mathrm{d}\psi =\frac{2}{\pi},
\label{eq:gap O2 c1}
\end{equation}
which means $c_1=1$.
Since the stability analysis focuses exclusively on the fundamental subharmonic mode, we approximate the tangent function~\eqref{eq:gap O2 fourier expansion} by its leading-order Fourier expansion term, which reads
\begin{equation}
\tanh \left ( \frac{\hat{\alpha}}{\epsilon} \left. \frac{\partial \eta_{p1}'}{\partial t'} \right |_{y'= b/2} \right )=\frac{2}{\pi} e^{\mathrm{i}\left ( \Omega t'+k_m x'+\pi/2 + \vartheta \left ( T' \right ) + \left. \vartheta _{\eta} \right |_{y'=b/2} \right ) }+\text{c.c.}
\label{eq:gap O2 tanh expression}
\end{equation}
Substituting~\eqref{eq:gap O2 tanh expression} into the expression of $\theta_{p2}$~\eqref{eq:gap O2 theta_2} and the contact line condition~\eqref{eq:gap O2 contact line condition}, one obtains
\begin{equation}
\frac{\partial \tilde{\eta} _{p2}'}{\partial y'} = -\frac{\mathrm{i} \hat{\rmDelta} }{\pi\sin^2 \theta_s }\frac{\check{\eta}_{p1}'}{\left | \check{\eta}_{p1}' \right | }\frac{B'}{\left | B' \right | },\quad \text{at }y'=\frac{b}{2}.
\label{eq:gap O2 final contact line condition}
\end{equation}
This condition gives a nonlinear correction to the previous order, which reflects the fast variation of contact angle in the hysteresis range at small contact line velocity.
Because the sgn function is utilized to approximate the hyperbolic tangent,~\eqref{eq:gap O2 final contact line condition} only depends on the phase of the velocity.
As the free surface fluctuates up-and-down, the contact line condition provides a different correction, which is consistent with the periodic feature of Faraday waves, and cannot be incorporated by the linear Hamraoui's model~\eqref{Hamraoui boundary}.

Substituting~\eqref{eq:gap O2 final contact line condition} into the solvability condition~\eqref{eq:gap O2 final compatibility condition}, an amplitude equation is obtained, leading to~\eqref{eq:gap amplitude equation}.

\end{appen}%\clearpage

\bibliographystyle{Manuscript}
\bibliography{Manuscript}

\end{document}